\documentclass[11pt]{article}
\usepackage[colorlinks]{hyperref}
\usepackage{amsmath}
\usepackage{amsthm}
\usepackage{thmtools}
\usepackage{amssymb}
\usepackage{graphicx}
\usepackage{algorithm}
\usepackage{algpseudocode}
\makeatletter
\renewcommand{\theHALG@line}{\thealgorithm.\arabic{ALG@line}}
\makeatother
\usepackage{multicol}
\usepackage{multirow}
\usepackage{color}
\usepackage{tikz}
\usepackage{xifthen}
\usetikzlibrary{decorations.pathreplacing,arrows.meta}

\usepackage[letterpaper,margin=1in,bottom=1in]{geometry}
\usepackage{etoolbox}
\patchcmd{\abstract}{\small}{\normalsize}{}{}
\usepackage[capitalize]{cleveref}
\usepackage[utf8]{inputenc}
\usepackage[english]{babel}
\usepackage{mathtools}

\usepackage{interval}
\intervalconfig{soft open fences}

\newtheorem{theorem}{Theorem}[section]

\newtheorem{lemma}[theorem]{Lemma}
\newtheorem{corollary}[theorem]{Corollary}
\newtheorem{proposition}[theorem]{Proposition}

\newtheorem{definition}[theorem]{Definition}

\newtheorem{remark}{Remark}[section]

\usepackage{tabularx}
\usepackage{booktabs}
\usepackage{threeparttable}
\usepackage{adjustbox}

\newcommand{\WS}{\textnormal{\textsc{WS}}}
\newcommand{\MT}{\textnormal{\textsc{MT}}}
\newcommand{\Prob}{\mathbb{P}}

\newcommand{\bits}{\{0,1\}}
\newcommand{\vars}{\operatorname{var}}

\newcommand{\Bad}{\operatorname{Bad}}

\newcommand{\one}{\mathbf{1}}

\newcommand{\Ftwo}{\mathbb F_2}

\newcommand{\eexp}{\mathrm e}

\newcommand{\Unif}{\operatorname{Unif}}

\newcommand{\E}[2][]{ \ifthenelse{\isempty{#1}}
  {\mathbf{E}\left[#2\right]}
  {\mathbf{E}_{#1}\left[#2\right]} }

\usepackage{authblk}

\title{Tight Variable-Occurrence Bounds and Exponential Separations for Moser--Tardos and WalkSAT on 3-CNF}

\author[1,2,3]{Minbo Gao}
\author[4]{Kun He}
\author[1,2,3]{Tao Jiang}

\affil[1]{Key Laboratory of System Software
(Chinese Academy of Sciences), Beijing, China}

\affil[2]{Institute of Software,
Chinese Academy of Sciences, Beijing, China}

\affil[3]{University of Chinese Academy of Sciences,
Beijing, China}

\affil[4]{Renmin University of China}

\date{\vspace{-1.2em}
\normalsize
\texttt{gaomb@ios.ac.cn},
\texttt{hekun2023@ruc.edu.cn},
\texttt{jiangt@ios.ac.cn}
}

\begin{document}

\maketitle

\begin{abstract}
We establish a sharp variable-occurrence threshold for polynomial expected-time convergence for the  Moser--Tardos algorithm
and the pure random-walk WalkSAT on 3-CNF formulas.
For every formula with $m$ clauses and at most two occurrences per variable, both algorithms
terminate in $O(m)$ expected updates.
This occurrence bound is sharp: we construct an infinite family
with at most three occurrences per variable on which both
algorithms require $2^{\Theta(\sqrt{m})}$ expected updates.
Surprisingly, this dynamical transition occurs strictly below the classical NP-completeness threshold at variable-occurrence degree four.

We further establish exponential separations between the two
algorithms on satisfiable 3-CNF formulas.
On one family, WalkSAT terminates in $O(m^4)$ expected flips,
whereas MT requires $2^{\Omega(m)}$ expected resampling steps.
On another, MT terminates in $O(m)$ expected resampling steps,
whereas WalkSAT requires $2^{\Omega(m)}$ expected flips.
These constructions exploit the different ways in which
single-variable flips and simultaneous resampling propagate
local violations.
Our results show that
the algorithms' expected running times are incomparable even
up to polynomial overhead on the same inputs.
\end{abstract}

\newpage
\tableofcontents
\newpage

\section{Introduction}
\label{sec:intro}
The Moser--Tardos (MT) resampling algorithm is a landmark achievement
in randomized algorithms. By making the Lov\'asz Local Lemma (LLL)
constructive in the variable setting, it transforms a powerful
existence theorem into an efficient procedure for finding
combinatorial objects that avoid prescribed local
obstructions~\cite{moser2010constructive}. Its simple update rule and
witness-tree analysis provide a striking example of how local
randomized dynamics can yield global algorithmic guarantees.
For Boolean satisfiability, the algorithm starts from a random
assignment and repeatedly resamples all variables in an unsatisfied
clause until it finds a satisfying assignment.

A natural parameter governing this process is the maximum number of
occurrences of a variable. In this paper, every clause of a $k$-CNF
formula contains exactly $k$ distinct variables. We write
$\Delta(F)$ for the maximum variable-occurrence degree of a formula
$F$, counting positive and negative occurrences together.
Bounding this parameter limits the interaction between local
repairs, since changing a variable can affect only the clauses
containing it. Nevertheless, this restriction alone does not explain
how successive repairs interact or when they lead to rapid
convergence.

The standard LLL analysis provides a starting point. Under the
uniform product distribution, a clause of a $k$-CNF formula is
unsatisfied with probability $p=2^{-k}$. If each variable occurs at
most $\Delta$ times, then each clause shares variables with at most
$D \leq k(\Delta-1)$
other clauses. Consequently, the symmetric LLL condition
$ep(D+1)\leq 1$, together with the Moser--Tardos theorem, guarantees
efficient resampling whenever
\[
    e\,2^{-k}\bigl(k(\Delta-1)+1\bigr)\leq 1.
\]
This gives a sufficient occurrence bound with leading term
$2^k/(ek)$.
For a 3-CNF formula with maximum variable-occurrence degree
$\Delta$, each clause is unsatisfied with probability $p=1/8$
under the uniform product distribution, and the clause-dependency
graph has maximum degree at most $D\le 3(\Delta-1)$.
Although the simplified symmetric LLL condition $epD\le 1$
already provides a sufficient occurrence bound, a sharper
conclusion follows from Shearer's criterion~\cite{shearer1985problem}.
In particular, if $\Delta\le 2$, then $D\le 3$ and
\[
    \frac18 < \frac{(3-1)^{3-1}}{3^3}=\frac4{27}.
\]
Together with the algorithmic extension of Kolipaka and
Szegedy~\cite{kolipaka2011moser}, this yields a linear expected
number of MT resampling steps.
However, this general dependency-degree analysis does not
provide a convergence guarantee for the entire degree-three
class, where the bound on the dependency degree increases to six.

The importance of exploiting additional structure is already
apparent in the asymptotic theory of bounded-occurrence $k$-SAT.
Let $f(k)$ be the largest integer $d$ such that every $k$-CNF formula
with maximum variable-occurrence degree at most $d$ is satisfiable.
Gebauer, Szab\'o, and Tardos established the asymptotically sharp
estimate
\[
    f(k)=
    \left(\frac{2}{e}
    +O\!\left(\frac{1}{\sqrt{k}}\right)\right)\frac{2^k}{k},
\]
improving the leading constant of the elementary symmetric-LLL
bound by a factor of two~\cite{gebauer2016local}.
Their lower bound uses the lopsided LLL and admits an algorithmic
implementation through the Moser--Tardos framework.
Crucially, this implementation uses a suitably biased,
formula-dependent product distribution. It must therefore be
distinguished from the particular MT process that initializes
and resamples every variable uniformly. Furthermore, asymptotic
tightness as $k\to\infty$ does not determine the exact behavior
at $k=3$.

Subsequent work further demonstrates that MT convergence can depend
on information beyond that captured by standard LLL criteria.
In particular, Harris developed a convergence criterion that
exploits the variable-level structure of bad events and can
establish efficient resampling beyond the lopsided LLL
region~\cite{harris2016lopsidependency}.
He, Li, and Sun~\cite{he2023beyond} subsequently used intersections
between dependent bad events to establish efficient MT convergence
beyond Shearer's bound for nonchordal dependency graphs.
However, such sufficient convergence criteria do not identify
the exact occurrence threshold for unbiased MT on 3-CNFs, which also
requires slow-instance lower bounds above the threshold.
These developments motivate studying the actual dynamics of MT
on bounded-occurrence 3-CNFs, rather than treating a general-purpose
LLL condition as its definitive algorithmic boundary.

A closely related perspective comes from stochastic local search.
WalkSAT and its variants are influential examples of clause-driven
search algorithms for satisfiability~\cite{selman1994noise}.
Although practical variants combine random choices with greedy
heuristics, we consider the pure random-walk version:
at each step, the algorithm selects an unsatisfied clause uniformly
at random and flips a uniformly random variable in that clause.
We refer to this process simply as WalkSAT throughout the paper.

Despite its simple description, the theoretical behavior of this
random walk remains only partially understood.
Papadimitriou proved that it finds a satisfying assignment of every
satisfiable 2-CNF formula in $O(n^2)$ expected
flips~\cite{papadimitriou1991selecting}.
More recently, WalkSAT was shown to run in linear expected time on typical random 2-SAT instances at every fixed clause density below the satisfiability threshold~\cite{berenbrink2025walksat}.
More precisely, with high probability over the input formula,
the expected number of flips, conditional on that formula,
is $O(n)$.

For random 3-CNFs, Alekhnovich and Ben-Sasson proved linear-time
guarantees below clause density
$1.63$~\cite{alekhnovich2007linear}.
For general $k$, a smoothed-analysis framework yields polynomial-time success with high probability up to densities of order $2^k/k^2$~\cite{cojaoghlan2009smoothed}.
Coja-Oghlan and Frieze subsequently proved linear-time success
up to density $\rho 2^k/k$, for an absolute constant $\rho>0$
and sufficiently large $k$~\cite{cojaoghlan2014analyzing}.
In the other direction, Coja-Oghlan, Haqshenas, and Hetterich showed
that at densities exceeding $c2^k\ln^2 k/k$, WalkSAT typically
has exponentially small success probability even after
exponentially many flips~\cite{cojaoghlan2017stalls}.
These results identify substantial regions of efficient and
inefficient behavior on random inputs, but do not provide
worst-case guarantees for deterministic bounded-occurrence formulas.

Bounded-occurrence 3-CNF provides a particularly sharp setting
in which to investigate this gap.
Tovey proved that every 3-CNF formula with maximum
variable-occurrence degree at most three is satisfiable,
whereas deciding satisfiability becomes NP-complete when the
degree bound increases to four~\cite{tovey1984simplified}.
The degree-three class also admits a deterministic matching-based
construction of a satisfying assignment.
Its tractability therefore leaves a separate question about
whether particular stochastic search processes find such an
assignment efficiently.

The NP-completeness of degree-four 3-CNF presents a
complexity-theoretic barrier to universal polynomial expected-time
guarantees. Indeed, a uniform polynomial expected-time bound for
either algorithm on all satisfiable instances in this class would
imply $\mathrm{NP}=\mathrm{RP}$: truncating the algorithm after a
suitable polynomial number of steps and verifying any returned
assignment would yield an RP algorithm for an NP-complete problem.
This argument, however, does not establish an exponential lower
bound. Determining a sharp threshold therefore requires matching
algorithm-specific results: a polynomial expected-time guarantee
for all satisfiable instances up to a given occurrence degree,
and a family of satisfiable instances with superpolynomially expected
running time at the next degree.

MT and WalkSAT approach these questions through closely related
local dynamics. Both repeatedly select an unsatisfied clause and
modify only its variables. Their update rules, however, differ
substantially. WalkSAT flips exactly one variable, whereas unbiased
MT independently redraws all three variables. Equivalently, an MT
update flips each selected variable independently with probability
$1/2$, and may therefore change any subset of the clause's variables,
including the empty set. Both updates can satisfy the selected
clause while making neighboring clauses unsatisfied.
Their relative performance consequently depends on how these
different repair rules propagate and resolve conflicts.

We study the two algorithms under a common probabilistic model.
Both start from a uniformly random assignment, select an
unsatisfied clause uniformly at each step, and run without
restarts. We measure running time by the number of local updates:
variable flips for WalkSAT and clause-resampling steps for MT.
Expectations are taken over initialization and all subsequent
random choices. Runtime guarantees and lower-bound constructions
concern satisfiable formulas.
Within this model, the preceding discussion highlights three closely related questions concerning the theoretical boundaries of WalkSAT and the MT algorithm and the relationship between them. For bounded-occurrence 3-CNF instances, these questions can be formulated as follows:
\begin{enumerate}
    \item \textbf{The MT Threshold on 3-CNF:} What is the exact variable occurrence threshold that guarantees the MT algorithm will find a solution in expected polynomial time for 3-CNF, and beyond what threshold does its expected running time inherently transition to superpolynomial?

    \item \textbf{The WalkSAT Threshold on 3-CNF:} Similarly, what is the critical threshold for the number of variable occurrences in 3-CNF instances below which WalkSAT is theoretically guaranteed to find a satisfying assignment in expected polynomial time, and above which there exist instances that require an expected superpolynomial running time?
    
    \item \textbf{The Computational Relationship on 3-CNF:} What is the structural and computational relationship between WalkSAT and the MT algorithm on 3-CNF? Specifically, can one algorithm theoretically simulate the other with only polynomial overhead, or is one strictly more powerful in terms of the class of instances it can efficiently solve?
\end{enumerate}

\subsection{Our results}
In this paper, we resolve these questions. We first show that, for every 3-CNF instance in which each variable occurs at most twice, both the WalkSAT and the MT algorithm have linear expected running time.
In contrast, allowing each variable to occur up to three times
already admits an infinite family on which both algorithms have
expected running time $2^{\Theta(\sqrt{m})}$.
Furthermore, we construct families of 3-CNF instances on which WalkSAT has exponential expected running time while the MT algorithm runs in expected polynomial time, as well as families exhibiting the opposite separation, where the MT algorithm has exponential expected running time while WalkSAT runs in expected polynomial time. These results show that the expected update complexities of
the two algorithms are incomparable up to polynomial overhead
on identical inputs, under the initialization and selection
rules specified above.

We consider 3-CNF formulas \(F=\bigwedge_{i=1}^{m}C_i\) over a set \(V=\{x_1,\ldots,x_n\}\) of Boolean variables, where each clause \(C_i\) is a disjunction of three literal occurrences,  involving distinct variables. 
An assignment satisfies \(F\) if every clause contains at least one true literal; a clause whose three literal occurrences are all false is called unsatisfied. 
For each variable \(x\in V\), let \(\Delta(x,F)\) denote the total number of occurrences of \(x\) and \(\neg x\) in \(F\), and define
$\Delta(F)\triangleq \max_{x\in V}\Delta(x,F)$.

Both WalkSAT and the MT algorithm follow clause-driven local-repair procedures, but their implementations admit several choices concerning initialization and the selection of clauses or variables. 
To facilitate a direct comparison, throughout this paper we adopt the following conventions. Initially, all distinct variables are assigned independently and uniformly at random. At each step, both algorithms select uniformly at random one of the currently unsatisfied clauses. WalkSAT then selects uniformly at random one of the three literal positions in the chosen clause and flips the underlying variable. By contrast, MT resamples independently and uniformly all variables appearing in the chosen clause. 
Both algorithms terminate when no unsatisfied clause remains.

These conventions are adopted only to specify canonical versions of the two algorithms and to streamline their comparison; they are not essential restrictions of our analysis. Our results extend to the broader classes of initialization and clause-selection rules specified in the corresponding theorem statements. We nevertheless formulate the algorithms under the above conventions in order to avoid unnecessary case distinctions and notation.

For a 3-CNF formula $F$, let $\E{T_{\MT}(F)}$ and $\E{T_{\WS}(F)}$ denote the expected running times of the MT algorithm and WalkSAT on $F$, respectively. When $F$ is clear from the context, we abbreviate these quantities as $\E{T_{\MT}}$ and $\E{T_{\WS}}$, respectively.

\vspace{0.2cm}

For any 3-CNF formula $F$ with $\Delta(F) \leq 2$, we have the following result.

\begin{theorem}\label{thm-bounded-two}
For every 3-CNF formula $F$ with $m$ clauses and $\Delta(F)\le2$, we have
\begin{align*}
\E{T_{\MT}(F)}\le m/2,\qquad \qquad \qquad \E{T_{\WS}(F)}\le 4m.
\end{align*}
\end{theorem}

On the contrary, WalkSAT and the MT algorithm cannot find solutions for all 3-CNF formulas with $\Delta(F) \leq 3$ in expected polynomial time.

\begin{theorem}
\label{thm-bounded-three}
There exists an infinite family of 3-CNF formulas
\(\{F_L\}_{L\geq 40}\), where \(F_L\) has $m=\Theta(L^2)$ clauses, $\Delta(F_L)\le 3$ such that
\begin{align*}
2^{\sqrt m/32}
 \le \E{T_{\MT}(F_L)}
 \le 2^{3\sqrt m},\qquad \qquad 2^{\sqrt m/32}
 \le \E{T_{\WS}(F_L)}
 \le 2^{3\sqrt m}.
\end{align*}
\end{theorem}

Although deciding the satisfiability of a general 3-CNF formula is NP-complete, Tovey showed that every 3-CNF formula in which each clause contains three variables and each variable occurs at most three times is satisfiable
~\cite{tovey1984simplified}. Moreover, a satisfying assignment can be constructed efficiently by finding a matching that saturates all clause vertices in the clause--variable incidence graph. The existence of such a matching follows immediately from Hall's theorem: every set of clauses has at least as many neighboring variables as clauses. Once the matching is found, each matched variable can be assigned the value that makes its occurrence in the corresponding clause true, thereby satisfying all clauses simultaneously. 
For a formula with \(m\) clauses, the clause-variable incidence graph
has at most \(3m\) edges and, after variables not occurring in the
formula are discarded, at most \(4m\) vertices.
Hence, the Hopcroft--Karp algorithm constructs such an assignment in $O(m^{3/2})$ time. In sharp contrast, the theorem above shows that WalkSAT and the MT algorithm may both require superpolynomial expected time even on formulas satisfying $\Delta(F)\leq 3$, revealing a fundamental limitation of these randomized local-repair algorithms.

Theorems \ref{thm-bounded-two} and \ref{thm-bounded-three} reveal a sharp transition in the worst-case performance of these algorithms as a function of the maximum variable occurrence: if every variable occurs at most twice, then both WalkSAT and the MT algorithm have linear expected running time on every 3-CNF formula; in contrast, allowing each variable to occur up to three times already admits instances on which either algorithm can have superpolynomial expected running time.
Surprisingly, this dynamical transition differs from the classical complexity transition for bounded-occurrence 3-CNF formulas.
A satisfying assignment can be found in polynomial time when each variable occurs at most three times, whereas deciding satisfiability becomes NP-complete when the occurrence bound increases to four~\cite{tovey1984simplified}.
This separation suggests that even 3-CNF formulas with at most three occurrences per variable can possess combinatorial structure that these algorithms fail to exploit efficiently.

\vspace{0.2cm}

To further separate the performance of WalkSAT and the MT algorithm on 3-CNF, we construct a family of satisfiable 3-CNF formulas on which WalkSAT has polynomial expected running time, whereas the MT algorithm requires exponentially many steps in expectation.

\begin{theorem}
\label{thm-mt-exponential}
There exists an integer $m_0$ and an infinite family of 3-CNF formulas
\(\{F_m\}_{m\geq m_0}\), where \(F_m\) has \(m\) clauses, and every clause contains exactly three distinct
variables, such that
\[
   \E{T_{\WS}}=O(m^4),
  \qquad \qquad \qquad
   \E{T_{\MT}}=2^{\Omega(m)}.
\]
\end{theorem}

On the contrary, we also construct an infinite family of satisfiable 3-CNF formulas on
which WalkSAT requires exponential expected time, whereas the MT algorithm terminates in linear time.

\begin{theorem}
\label{thm-ws-exponential}
There exists an infinite family of satisfiable 3-CNF formulas
$\{F_L\}_{L\geq 1}$ such that, writing $m_L$ for the number of
clauses of $F_L$, we have $m_L\to\infty$ as $L\to\infty$, and
\[
  \E{T_{\WS}(F_L)}=2^{\Omega(m_L)},
  \qquad
  \E{T_{\MT}(F_L)}=O(m_L).
\]
\end{theorem}

The two separating families show that the algorithms' expected
running times are incomparable even up to polynomial overhead
on the same inputs. Thus, their closely related local-repair
paradigms support fundamentally different convergence behavior:
each algorithm can efficiently solve instances on which the
other requires exponential expected time.

\subsection{Technique overview}

Both WalkSAT (\WS) and Moser--Tardos (\MT) repeatedly select an unsatisfied clause, but differ in how they update its variables. \WS\ flips a single variable, which immediately satisfies the selected clause, whereas \MT\ independently resamples all distinct variables in the clause, potentially changing several variables simultaneously. This difference has two competing effects: resampling may help satisfy multiple constraints at once, but it may also introduce new violations. Our proofs exploit these properties to establish both efficient convergence and exponential lower bounds.

We first show that the occurrence threshold is sharp (\cref{sec:occurrence-two,sec:joint-hard-family}). When every variable occurs at most twice, the dependencies between clauses are sufficiently restricted for both algorithms to run in linear expected time. For \MT, we show that the cluster-expansion criterion in~\cite{Pegden2014} applies. For \WS, we represent the formula as a graph and use the component classification (\cref{lem:ws-component-classification}) to analyze defect trees and boundary trees. Writing $\Phi_t$ for the weighted potential at time $t$ (\cref{eq:ws-global-potential}), we show that, before termination,
\[
  \E{\Phi_{t+1}-\Phi_t\mid\mathcal F_t}
  \le -\frac14,
\]
giving a linear expected-time bound. In contrast, when variables may occur three times, we construct the formulas $F_L$ from chains and modules with long tails (\cref{subsec:proper-construction}). We show that every satisfying assignment must either satisfy certain conditions along an entire branch or set all auxiliary variables in some module to $1$ (\cref{lem:proper-terminal-structure}). With the branch factors $K_j$ and module factors $Q_{j,i}$ (\cref{eq:proper-chain-potential,eq:proper-module-potential}), we construct the product potential
\[
  P(X)\triangleq K_1(X)K_2(X)K_3(X)
       \prod_{j=1}^{3}\prod_{i=1}^{L-1}Q_{j,i}(X),
\]
which vanishes at every satisfying assignment. The key observation is that updates to non-root clauses cannot decrease the branch potentials in expectation, while the expected decrease of the product potential is limited. This yields $2^{\Omega(L)}$ lower bounds for both algorithms. Together with matching upper bounds and $m=\Theta(L^2)$, we obtain expected running times of $2^{\Theta(\sqrt m)}$.

To obtain a separation in favor of \WS, we consider the zero-charge Tseitin formulas $F(G)$ on $3$-regular graphs $G=(U,E)$ (\cref{subsec:tseitin-encoding}). Let $D_t=D(X_t)$ denote the set of vertices whose parity constraints are violated (\cref{eq:tseitin-defect-set}). When \WS\ flips an edge variable $x_e$ for $e=\{u,v\}$, where $v\in D_t$, we have
\[
  D_{t+1}=D_t\mathbin\triangle\{u,v\}.
\]
Thus, a \WS\ step either moves a violation to a neighboring vertex or eliminates two violations. In particular, \WS\ never increases the number of violated parity constraints, and its dynamics form an annihilating random walk with polynomial expected termination time. In contrast, \MT\ resamples all three edge variables incident to a violated vertex and may create new violations at its neighbors. We show that the continuous-time version of the resulting process is dual to the threshold voter model (\cref{lem:mt-fourier-duality}). On locally sparse $3$-regular graphs, we show that the associated threshold contact process survives for exponential time (\cref{lem:contact-survival}), which yields an exponential lower bound for \MT.

For the reverse separation, we construct formulas $F(G)$ from expander graphs $G=(V,E)$ (\cref{subsec:direct-construction}). Each vertex $v$ has two variables $p_v,q_v$ and is active exactly when $(p_v,q_v)=(1,1)$. Let $S_t=S(X_t)$ be the set of active vertices at time $t$. The local gadgets exploit \MT's ability to resample multiple variables simultaneously. For example, when $(p_v,q_v)=(0,0)$, resampling an unsatisfied clause $(p_v\lor q_v\lor h_{v,c})$ activates $v$ with probability $1/4$, whereas a single \WS\ flip cannot activate it. By balancing the contributions of the local gadgets and edge clauses, we show that, before termination, \MT\ has a uniform positive drift in the number of active vertices (\cref{lem:direct-mt-drift}):
\[
  \E{|S_{t+1}|-|S_t|\mid\mathcal F_t}
  \ge \frac{1}{3684},
\]
which implies linear expected running time. For \WS, however, repairing an edge clause may deactivate an already active vertex instead of activating its neighbor. In particular, an unsatisfied edge clause from an active vertex to a vertex with one zero variable increases the number of active vertices with probability $1/3$, but decreases it with probability $2/3$. Using the expansion property, we show that this unfavorable bias persists whenever $H\le |S_t|<|V|$, where $H=\lceil3|V|/4\rceil$ (\cref{lem:direct-ws-bias}):
\[
  p_t^+\le \rho p_t^-,
  \qquad \rho=\frac{451}{486}<1,
\]
where $p_t^\pm=\Prob\bigl(|S_{t+1}|-|S_t|=\pm1\mid\mathcal F_t\bigr)$. Consequently, we show that \WS\ has an exponentially small probability of reaching a satisfying assignment in a single attempt (\cref{prop:direct-ws-one-attempt}), leading to exponential expected running time.

These results demonstrate that single-variable flips and clause-wide resampling have different advantages. On Tseitin formulas, \WS\ avoids creating additional parity violations, while \MT\ may sustain them for exponential time. On the expander-based formulas, \MT\ can make progress by resampling multiple variables simultaneously, while \WS\ is hindered by updates that undo previous progress. Thus, neither update rule is uniformly more efficient than the other.

\subsection{Organization}

The remainder of this paper is organized as follows. \Cref{sec:model} introduces the notation, definitions, and preliminary results used throughout the paper.
\Cref{sec:occurrence-two} proves that, for every 3-CNF formula in which each variable occurs at most twice, both WalkSAT and the MT algorithm run in expected linear time. 
\Cref{sec:joint-hard-family} shows that allowing each variable to occur up to three times already admits 3-CNF instances on which both algorithms require superpolynomial expected time. 
\Cref{sec:mt-slow} constructs a family of 3-CNF instances on which WalkSAT has polynomial expected running time, whereas the MT algorithm has exponential expected running time. 
\Cref{sec-ws-slow} establishes the converse separation by constructing a family of 3-CNF instances on which the MT algorithm runs in expected linear time, whereas WalkSAT requires exponential expected time. 
Finally, \Cref{sec:conclusion} concludes the paper with a summary of the main results.

\section{Preliminary}
\label{sec:model}

\subsection{Formula representation and occurrence counts}


We consider 3-CNF formulas \(F=\bigwedge_{i=1}^{m}C_i\) over a set \(V=\{x_1,\ldots,x_n\}\) of Boolean variables, where each clause \(C_i\) is a disjunction of three literal occurrences, involving distinct variables. 
For an assignment $\sigma\in\bits^V$, let $\Bad_F(\sigma)$ be the set of
indices of clauses falsified by $\sigma$.  The hitting time of the satisfying
set is denoted by
\[
   T=\inf\{t\ge 0:\Bad_F(X_t)=\varnothing\}.
\]
We count resampled clauses for Moser--Tardos and flipped variables for
WalkSAT.

\subsection{The two algorithms and clause-selection rules}

Both algorithms maintain a complete assignment.  At each time $t<T$, a
clause-selection rule chooses an index in $\Bad_F(X_t)$.  An adaptive rule may
depend on the history through time $t$ but not on future random choices.  The
uniform rule chooses each index in $\Bad_F(X_t)$ with probability
$1/|\Bad_F(X_t)|$.

\begin{definition}[WalkSAT]
Given an initial assignment $X_0$, select an unsatisfied clause according to
the stated rule, choose one of its three literal occurrences uniformly, and
flip the corresponding variable. We abbreviate the
process by $\WS$; see
\cite{papadimitriou1991selecting,Schoning1999} for the underlying random-walk method.
\end{definition}

\begin{algorithm}[H]
\caption{WalkSAT ($\WS$)}
\label{alg:randomwalksat}
\begin{algorithmic}
  \Require Indexed 3-CNF $F=(C_1,\ldots,C_m)$, initial assignment $X_0$,
    and a clause-selection rule
  \State $t\gets 0$
  \While{$\Bad_F(X_t)\ne\varnothing$}
    \State Choose $i_t\in\Bad_F(X_t)$ according to the clause-selection rule
    \State Choose a literal occurrence of $C_{i_t}$ uniformly at random
    \State Let $x_t$ be the variable underlying the chosen occurrence
    \State Set $X_{t+1}\gets X_t$ and
      $X_{t+1}(x_t)\gets 1-X_t(x_t)$
    \State $t\gets t+1$
  \EndWhile
  \State \Return $X_t$
\end{algorithmic}
\end{algorithm}

\begin{definition}[Moser--Tardos resampling algorithm]
Initially sample all variables independently and uniformly from $\bits$.
Select an unsatisfied clause $C$ according to the stated rule and replace each
variable in $\vars(C)$ by a fresh independent unbiased bit. We
abbreviate the process by $\MT$~\cite{moser2010constructive}.
\end{definition}

\begin{algorithm}[H]
\caption{Moser--Tardos resampling algorithm ($\MT$)}
\label{alg:moser-tardos}
\begin{algorithmic}
  \Require Indexed 3-CNF $F=(C_1,\ldots,C_m)$ and a clause-selection rule
  \State Independently sample $X_0(x)\sim\Unif(\bits)$ for every variable $x$
  \State $t\gets 0$
  \While{$\Bad_F(X_t)\ne\varnothing$}
    \State Choose $i_t\in\Bad_F(X_t)$ according to the clause-selection rule
    \State $X_{t+1}\gets X_t$
    \ForAll{$x\in\vars(C_{i_t})$}
      \State Independently sample $X_{t+1}(x)\sim\Unif(\bits)$
    \EndFor
    \State $t\gets t+1$
  \EndWhile
  \State \Return $X_t$
\end{algorithmic}
\end{algorithm}

Unless a theorem states otherwise, both algorithms start from the uniform
distribution on assignments and use the uniform clause-selection rule.

\subsection{Satisfiability at maximum variable occurrence three}

\begin{proposition}
\label{prop:hall}
Every formula in the representation above with $\Delta(F)\le3$ is
satisfiable.
\end{proposition}

\begin{proof}
Form the bipartite incidence graph with clauses on the left and variables on
the right.  Let $\mathcal S$ be any set of clauses.  Its clauses contain
$3|\mathcal S|$ literal occurrences, while every variable in their
neighborhood accounts for at most three of them.  Hence
$|N(\mathcal S)|\ge |\mathcal S|$.  Hall's theorem~\cite{Hall1935} gives a
matching that assigns a distinct variable to every clause.  Set each matched
variable so that one of its occurrences in its matched clause is true, and
set all unmatched variables arbitrarily.  The matched variables are distinct,
so these requirements are consistent and every clause is satisfied.
\end{proof}

The proposition fails if clauses with fewer than three literal occurrences
are allowed: $(x)\wedge(\neg x)$ is unsatisfiable even though $x$ occurs only
twice.

\subsection{Two drift lemmas}

The following two lemmas give the drift bounds used below.

\begin{lemma}
\label{lem:unit-drift}
Let $T$ be a hitting time and let $\Phi$ take values in a finite interval
$[a,b]$.  If for every $t<T$,
\[
  \E{\Phi(X_{t+1})-\Phi(X_t)\mid\mathcal F_t}=-1,
\]
then $T<\infty$ almost surely and
\[
  \E{T}\le \E{\Phi(X_0)}-a\le b-a.
\]
The same conclusion holds with ``$\le-1$'' in place of equality.
\end{lemma}

\begin{proof}
For deterministic $s$, the stopped process gives
$\E{s\wedge T}\le\E{\Phi(X_0)}-a$.  Letting $s$ tend to infinity and applying
monotone convergence proves both assertions.
\end{proof}

\begin{lemma}
\label{lem:drift-lower-bound}
Let $\tau$ be an integrable stopping time and let $0\le Z_t\le B$, with
$Z_\tau=0$.  Suppose that before $\tau$ the conditional expected loss per
step is at most $D>0$:
\[
  \E{Z_t-Z_{t+1}\mid\mathcal F_t}\le D.
\]
Then $\E{\tau}\ge \E{Z_0}/D$.  Consequently, if $\tau\le T$, then
$\E{T}\ge\E{Z_0}/D$.
\end{lemma}

\begin{proof}
Apply the drift inequality to $\tau\wedge s$ and telescope.  Dominated
convergence for $Z_{\tau\wedge s}$ and monotone convergence for
$\tau\wedge s$ yield $\E{Z_0}\le D\E{\tau}$.
\end{proof}

Before each application of Lemma~\ref{lem:drift-lower-bound}, we separately
prove that the relevant hitting time is integrable.

\section{At most two occurrences per variable}
\label{sec:occurrence-two}

In this section, we analyze the running times of MT and WalkSAT on 3-CNF formulas $F$ with $\Delta(F)\leq 2$.
The respective bounds, established in \Cref{thm:occurrence-two-mt,thm:occurrence-two-ws}, immediately imply \Cref{thm-bounded-two}.

\subsection{Moser--Tardos via the cluster-expansion criterion}

\begin{theorem}
\label{thm:occurrence-two-mt}
Let $F=(C_1,\ldots,C_m)$ be a 3-CNF formula with
that $\Delta(F)\le2$.  Starting from
the uniform product distribution, Moser--Tardos satisfies
\[
  \E{T_{\MT}}\le\frac m2\le\frac n3,
\]
where $n$ is the number of variables occurring in $F$.  The bound holds for
every adaptive rule for selecting an unsatisfied clause.
\end{theorem}

We recall Pegden's algorithmic cluster-expansion local
lemma~\cite{Pegden2014} as follows.

\begin{theorem}[{\cite[Theorem~1.4]{Pegden2014}}]
\label{thm:cluster-expansion-lll}
 Let $\mathcal{V}$ be a finite set of mutually independent variables, and  $\mathcal{A}$ be a finite family of events determined by these
variables.  Let $G$ be the graph on $\mathcal A$ in which two events are
adjacent if they depend on a common variable.  For $A\in\mathcal A$, let
$\bar{\Gamma}(A)$ be $A$ together with all its neighbors in $G$; that is,
\[
  \bar{\Gamma}(A)=\{A\}\cup\{B\in\mathcal A:\{A,B\}\in E(G)\}.
\]
If there are positive numbers $(\mu_A)_{A\in\mathcal A}$, such that
\[
  P(A)\le \frac{\mu_A}{\sum_{\substack{I\subseteq\bar{\Gamma}(A)\\
                   I\text{ indep.}}}
    \prod_{B\in I}\mu_B}
\]
for every $A\in\mathcal A$,
then the Moser--Tardos algorithm terminates almost surely under any rule that
selects a currently occurring bad event.  Moreover, if $N_A$ is the number of
times that $A$ is resampled, then
\[
  \E{N_A}\le\mu_A
  \quad\text{for every }A\in\mathcal A,
  \qquad
  \E{T_{\MT}}\le\sum_{A\in\mathcal A}\mu_A.
\]
\end{theorem}

\begin{proof}[Proof of Theorem~\ref{thm:occurrence-two-mt}]
Let $V$ be the set of variables occurring in $F$, let $\Omega=\bits^V$, and
equip $\Omega$ with the uniform product measure
\[
  \pi=\bigotimes_{x\in V}\Unif(\bits).
\]
For each indexed clause $C\in F$, define its bad event as the set
\[
  A_C=\{\sigma\in\Omega:C\text{ is false under }\sigma\}.
\]
Write
\[
  \Psi_{A_C}
  =\sum_{\substack{I\subseteq\bar{\Gamma}(A_C)\\I\text{ independent}}}
    \prod_{A_D\in I}\mu_{A_D}.
\]
Set $\mu_{A_C}=1/2$ for every clause $C$. Since $C$ contains three distinct variables, then
$\pi(A_C)=1/8$, and $A_C$ has at most three distinct neighbors.  An independent
set in $\bar{\Gamma}(A_C)$ either contains $A_C$ alone or is a subset of these
neighbors.  Consequently,
\[
  \Psi_{A_C}
  \le \frac12+\left(1+\frac12\right)^3
  =\frac{31}{8},
  \qquad
  \frac{\mu_{A_C}}{\Psi_{A_C}}
  \ge\frac4{31}>\frac18.
\]
Summing the per-event bounds proves
\[
  \E{T_{\MT}}\le\sum_{C\in F}\mu_{A_C}=\frac m2.
\]
The symmetric criterion based only on maximum dependency degree would require
$e(1/8)(3+1)\le1$, which does not hold~\cite{moser2010constructive}.
Finally, $3m\le2n$ because the formula has $3m$ literal occurrences and each
of its $n$ variables occurs at most twice.
\end{proof}

\subsection{WalkSAT via a global potential}

The graph representation below requires opposite signs at the two occurrences
of a shared variable: its edge must have exactly one true endpoint.  The
following proposition reduces the general case to this setting by showing
that splitting same-sign occurrences can only increase the termination time
under a suitable coupling.

\begin{proposition}[Same-sign splitting]
\label{prop:ws-splitting}
Let $\Delta(F)\le2$, and let $S$ be the set of variables with two same-sign
occurrences.  For each $x\in S$, introduce fresh variables $x^{(1)},x^{(2)}$
and replace its two literal occurrences by
\[
\begin{aligned}
  (x,x)&\longmapsto(x^{(1)},x^{(2)}),\\
  (\neg x,\neg x)&\longmapsto(\neg x^{(1)},\neg x^{(2)}).
\end{aligned}
\]
Call the resulting formula $F^\sharp$; all other literals are unchanged.
Given an initial assignment $X_0$ on $V$, define $Y_0$ by
\[
\begin{aligned}
  Y_0(x^{(1)})=Y_0(x^{(2)})&=X_0(x) &&(x\in S),\\
  Y_0(z)&=X_0(z) &&(z\in V\setminus S).
\end{aligned}
\]
Write $T_{\WS}(G;\sigma,\mathcal R)$ for the termination time on $G$ from
$\sigma$ under an adaptive clause-selection rule $\mathcal R$.  Then
\[
\begin{gathered}
  \forall X_0\in\bits^V,\quad \forall\mathcal R\text{ on }F,\quad
  \exists\mathcal R^\sharp\text{ on }F^\sharp\text{ and a coupling}:\\[2pt]
  \Prob\!\left(
    T_{\WS}(F;X_0,\mathcal R)
    \le T_{\WS}(F^\sharp;Y_0,\mathcal R^\sharp)
  \right)=1.
\end{gathered}
\]
\end{proposition}

\begin{proof}
Until $F$ stops, use its clause index and the same uniform literal position
in both walks.  Let $a_p(t),b_p(t)$ be the truth values at corresponding
positions under the assignments $X_t,Y_t$.  The invariant is
\[
  a_p(t)\ge b_p(t)\quad(\forall p),
  \qquad
  a_p(t)=b_p(t)\quad(p\text{ not split}).
\]
It holds initially and implies
\[
  \Bad_F(X_t)\subseteq\Bad_{F^\sharp}(Y_t),
\]
so every coupled clause choice is legal.  Flipping an unsplit variable
preserves equality.  Selecting a split variable at position $p$, with
other occurrence $q$, gives
\[
  (a_p,a_q;b_p,b_q)=(0,0;0,0)\longmapsto(1,1;1,0),
\]
which preserves the invariant.  Thus $F^\sharp$ cannot stop before $F$.
After $F$ stops, continue $F^\sharp$ if necessary.  The induced rule uses
no future randomness and is adaptive, though not necessarily uniform.
\end{proof}

\begin{theorem}
\label{thm:occurrence-two-ws}
Let $F=(C_1,\ldots,C_m)$ be a 3-CNF formula with $\Delta(F)\le2$.  
WalkSAT
satisfies
\[
  \E{T_{\WS}}\le4m\le\frac83n
\]
from every initial assignment, where $n$ is the number of variables occurring
in $F$.  The bound holds for every adaptive rule for selecting an unsatisfied
clause.
\end{theorem}

\begin{proof}
We first consider the case in which every variable occurring twice has one
positive and one negative occurrence. Since each clause contains three distinct variables, these two occurrences lie in different clauses. Every other variable occurs
once.

Form a multigraph with half-edges whose vertices are the clauses.  The two
oppositely signed occurrences of a variable give an edge between their
clauses, and each unique occurrence gives a half-edge.  Every clause vertex
has three incident positions.

For the current assignment, orient every edge toward its true literal.  Thus
each edge has exactly one true endpoint.  Let $s(C)$ be the number of true
literal occurrences in $C$, and define
\[
  A=\{C:s(C)\ge2\}.
\]
Delete the vertices in $A$ and all incident edges.  A half-edge or a deleted
edge incident to a remaining component is a \emph{boundary occurrence}.

\begin{lemma}
\label{lem:ws-component-classification}
Every connected component outside $A$ is exactly one of the following:
\begin{enumerate}
  \item a tree with one unsatisfied clause and no true boundary occurrence;
  \item a tree with no unsatisfied clause and one true boundary occurrence;
  \item a unicyclic component with no unsatisfied clause and no true boundary
    occurrence.
\end{enumerate}
\end{lemma}

\begin{proof}
Let a component $K$ have $q$ vertices and $e$ edges.  Let $u$ be its number of
unsatisfied clauses and $b$ its number of true boundary occurrences.  Every
vertex outside $A$ has zero or one true occurrence, so
\[
  e+b=\sum_{C\in K}s(C)=q-u.
\]
Connectivity gives $e\ge q-1$.  Hence $u\le1$.  If $u=1$, then
$e=q-1$ and $b=0$.  If $u=0$, either $e=q-1$ and $b=1$, or $e=q$ and $b=0$.
These are precisely the three stated cases.
\end{proof}

Call the first type a \emph{defect tree}, rooted at its unsatisfied clause,
and the second a \emph{boundary tree}, rooted at the clause carrying its
unique true boundary occurrence.  For a rooted tree $(T,r)$, set
\[
  W_T(r)=\sum_{v\in V(T)}2^{-d_T(r,v)}.
\]
Define the global potential
\begin{equation}
  \Phi
  =\sum_{\substack{T\text{ defect tree}\\r_T\text{ defect root}}}W_T(r_T)
  +\frac14
   \sum_{\substack{T\text{ boundary tree}\\\rho_T\text{ boundary root}}}
      W_T(\rho_T).
  \label{eq:ws-global-potential}
\end{equation}
Unicyclic components and vertices in $A$ contribute zero.  Since
$W_T(r)\le |V(T)|$ and the components are disjoint,
\begin{equation}
  0\le\Phi\le m.
  \label{eq:ws-global-potential-bound}
\end{equation}

We need two estimates for the change in this potential.  First, if the root
$r$ of a defect tree $T$ has degree $d$ in $T$, then
\begin{equation}
  \sum_{u\sim_T r}W_T(u)=\frac{d+3}{2}W_T(r)-\frac32.
  \label{eq:ws-rerooting-identity}
\end{equation}
Indeed, after deleting $r$, let $a_u$ be the contribution to $W_T(r)$ of the
branch containing $u$.  Then $W_T(r)=1+\sum_{u\sim r}a_u$, while
\[
  W_T(u)=2a_u+\frac12
    +\frac12\sum_{\substack{v\sim r\\v\ne u}}a_v.
\]
Summing over $u$ proves~\eqref{eq:ws-rerooting-identity}.

Second, consider choosing at $r$ an occurrence that is not an edge of $T$.
Write $W=W_T(r)$ and let $Q$ be the contribution to $\Phi$ of all other
components before the update.  Then
\begin{equation}
  \Phi_{t+1}\le Q+\frac14W+\frac12.
  \label{eq:ws-boundary-update}
\end{equation}
If the occurrence is unique, the defect tree becomes a boundary tree and
contributes $W/4$.  The same holds if the occurrence leads to a clause
$D\in A$ that remains in $A$ after the flip.  In both cases every other
component retains its contribution to $Q$.

It remains to justify~\eqref{eq:ws-boundary-update} when $s(D)$ decreases from
two to one.  Before the flip the selected edge points from $r$ to $D$;
afterwards it points from $D$ to $r$.  Call the other two positions of $D$
$\alpha$ and $\beta$, where $\alpha$ is true at $D$ and $\beta$ is false at
$D$.  Their truth values do not change.  If these positions have neighbors
outside the old set $A$, their orientations are $b\to D$ and $D\to c$,
respectively (\cref{fig:ws-boundary-surgery}).  In particular, the occurrence
at $c$ is a true boundary occurrence before the update.  The component
classification therefore forces its component to be a boundary tree $L$
rooted at $c$.  Thus $\beta$ cannot attach to $T$, another defect tree, or a
unicyclic component.  At $b$ the boundary occurrence is false, so $\alpha$
may attach to any of the three component types, including $T$ itself.
An attachment is called absent, and denoted by $\varnothing$, if the position
is a half-edge or its other endpoint remains in $A$.

\begin{figure}[htbp]
\centering
\begin{tikzpicture}[>=Stealth,
  clause/.style={circle,draw,inner sep=2pt,minimum size=6mm},
  every node/.style={font=\normalsize}]
  \begin{scope}
    \draw[gray] (0,0) ellipse (0.65 and 0.45);
    \draw[gray] (2,1.45) ellipse (0.65 and 0.45);
    \draw[gray] (4.1,0) ellipse (0.65 and 0.45);
    \node at (-0.55,0.65) {$T$};
    \node at (2,2.05) {$K$};
    \node at (4.65,0.65) {$L$};
    \node[clause] (r0) at (0,0) {$r$};
    \node[clause,fill=gray!20] (D0) at (2,0) {$D$};
    \node[clause] (b0) at (2,1.45) {$b$};
    \node[clause] (c0) at (4.1,0) {$c$};
    \draw[->,dashed] (r0) -- node[above] {selected} (D0);
    \draw[->,dashed] (b0) -- node[right] {$\alpha$} (D0);
    \draw[->,dashed] (D0) -- node[above] {$\beta$} (c0);
    \node at (0,-0.7) {$s(r)=0$};
    \node at (2,-0.7) {$s(D)=2$};
    \node[align=center] at (4.1,-0.8)
      {true boundary\\at $c$};
    \node at (2,-1.45) {(a) Before: $D\in A$};
  \end{scope}
  \begin{scope}[xshift=7cm]
    \draw[gray] (0,0) ellipse (0.65 and 0.45);
    \draw[gray] (2,1.45) ellipse (0.65 and 0.45);
    \draw[gray] (4.1,0) ellipse (0.65 and 0.45);
    \node at (-0.55,0.65) {$T$};
    \node at (2,2.05) {$K$};
    \node at (4.65,0.65) {$L$};
    \node[clause] (r1) at (0,0) {$r$};
    \node[clause] (D1) at (2,0) {$D$};
    \node[clause] (b1) at (2,1.45) {$b$};
    \node[clause] (c1) at (4.1,0) {$c$};
    \draw[->] (D1) -- node[above] {selected} (r1);
    \draw[->] (b1) -- node[right] {$\alpha$} (D1);
    \draw[->] (D1) -- node[above] {$\beta$} (c1);
    \node at (0,-0.7) {$s(r)=1$};
    \node at (2,-0.7) {$s(D)=1$};
    \node at (2,-1.45) {(b) After: $D\notin A$};
  \end{scope}
\end{tikzpicture}
\caption{The surgery when $D$ leaves $A$, illustrated with distinct attached
components $T,K,L$.  Arrows point toward true occurrences.  Before the flip,
the dashed edges are deleted with $D$: $D\to c$ is true at the boundary of
$L$, whereas $b\to D$ is false at the boundary of $K$.  Only the selected
edge reverses.  Absent and coincident attachments are included in
\cref{tab:ws-boundary-attachments}.}
\label{fig:ws-boundary-surgery}
\end{figure}
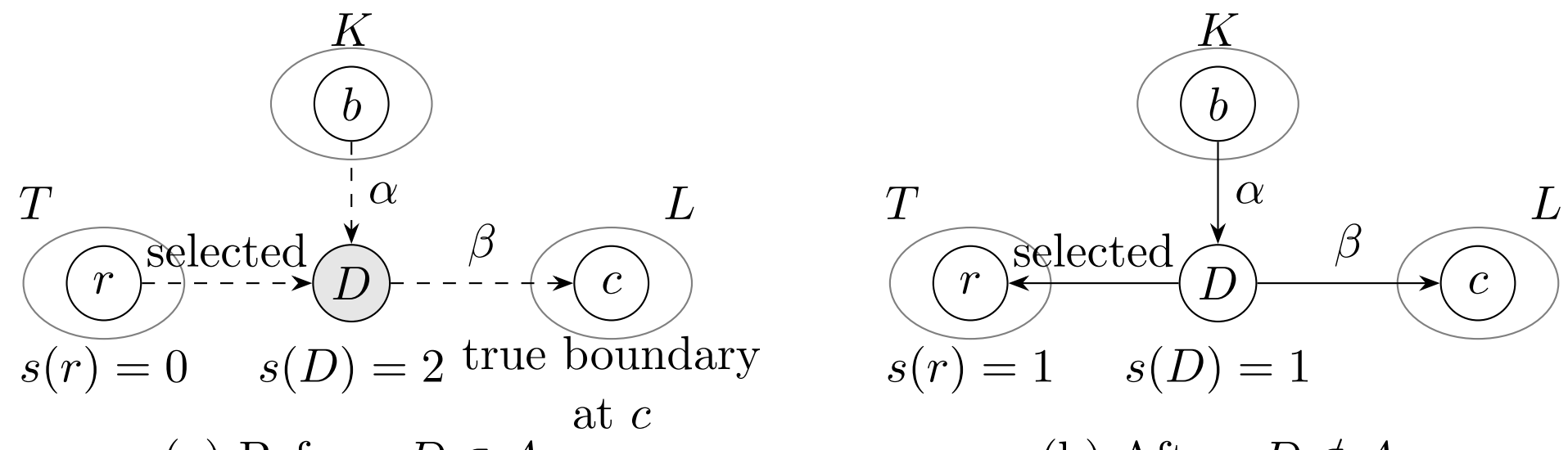

Only $r$ and $D$ change their numbers of true occurrences, and $r$ remains
outside $A$.  Consequently, the new graph outside $A$ is obtained solely by
adding $D$ and its incident edges to the old components.  No old component
splits.  Let $\mathcal C$ be the set of distinct old components other than
$T$ touched by $\alpha$ or $\beta$, and let $\varphi(K)$ denote the
contribution of a component $K$ to~\eqref{eq:ws-global-potential}.  All
components in $\mathcal C$ merge with $T$ and $D$ into one component, whose
contribution we denote by $\varphi_{\mathrm{new}}$.  Every untouched
component has exactly the same contribution as before.  Hence, with
\[
  \Delta=\varphi_{\mathrm{new}}
          -\sum_{K\in\mathcal C}\varphi(K),
  \qquad
  \Phi_{t+1}=Q+\Delta,
\]
it suffices to prove $\Delta\le W/4+1/2$.  Each old component is subtracted
only once, even if both positions attach to it.

\begin{table}[htbp]
\centering
\normalsize
\renewcommand{\arraystretch}{1.2}
\begin{tabularx}{\textwidth}{@{}p{0.25\textwidth}p{0.14\textwidth}>{\raggedright\arraybackslash}Xp{0.19\textwidth}@{}}
\toprule
At $\alpha$ & At $\beta$
  & New component & Bound on $\Delta$\\
\midrule
$\varnothing$ & $\varnothing$ or $L$
  & Boundary tree rooted at $D$ & $\frac14+\frac18W$\\
Defect tree $S\ne T$ & $\varnothing$ or $L$
  & Defect tree rooted at the old root $x$ of $S$
  & $\frac12+\frac14W$\\
Boundary tree $B\ne L$ & $\varnothing$ or $L$
  & Boundary tree rooted at the old root $z$ of $B$
  & $\frac18+\frac1{16}W$\\
$T$ & $\varnothing$ or $L$
  & Unicyclic: the two edges from $D$ to $T$ close a cycle & $0$\\
Boundary tree $L$ & $L$
  & Unicyclic: $\alpha$ and $\beta$ enter the same tree & $0$\\
Unicyclic component & $\varnothing$ or $L$
  & Unicyclic: the existing cycle is retained & $0$\\
\bottomrule
\end{tabularx}
\caption{All attachment cases when $s(D)$ drops from $2$ to $1$.
The condition $B\ne L$ is vacuous when the attachment at $\beta$ is absent.}
\label{tab:ws-boundary-attachments}
\end{table}

The cases in \cref{tab:ws-boundary-attachments} are exhaustive: $\beta$ has
only the two possibilities already identified, and $\alpha$ is either
absent or enters $T$, a different defect tree, a boundary tree, or a
unicyclic component.  The only possible coincidence between its component
and that of $\beta$ is the row in which both enter $L$.

For the three acyclic cases, put $\lambda=W_L(c)$ when $L$ is present and
$\lambda=0$ otherwise.  In the first row, $\alpha$ supplies the unique true
boundary occurrence at $D$, so
\[
  \Delta
  =\frac14\left(1+\frac12W+\frac12\lambda\right)
     -\frac14\lambda
  =\frac14+\frac18W-\frac18\lambda
  \le\frac14+\frac18W.
\]
In the second row, let $b\in S$ be the attachment vertex and set
$\delta=d_S(x,b)$.  Distances from $x$ to $D$ and $r$ are $\delta+1$ and
$\delta+2$; when $L$ is present, the distance to $c$ is also $\delta+2$.
Thus the new weight
at $x$ equals the old weight $W_S(x)$ plus
$2^{-\delta}(1/2+W/4+\lambda/4)$.  The term $W_S(x)$ cancels against its
old contribution in $Q$, leaving
\[
  \Delta
  =2^{-\delta}\left(\frac12+\frac14W+\frac14\lambda\right)
     -\frac14\lambda
  \le\frac12+\frac14W.
\]
Here $\delta\ge0$, so the coefficient of $\lambda$ is nonpositive.  In the
third row the same distance calculation applies from $z$, with
$\delta=d_B(z,b)$, but the new tree and $B$ both have coefficient $1/4$ in
the potential.  After canceling the old contribution of $B$, we obtain
\[
  \Delta
  =\frac14\,2^{-\delta}
      \left(\frac12+\frac14W+\frac14\lambda\right)
     -\frac14\lambda
  \le\frac18+\frac1{16}W.
\]
Again the contribution of $L$ is fully covered by its subtraction from $Q$;
when the attachment is absent, the same calculations hold with $\lambda=0$.

In the last three rows, the merged component is unicyclic.  In particular,
two attachments to the same tree create one cycle, including a cycle of
length two if the two edges are parallel.  Attaching to an old unicyclic
component only adds trees to its existing cycle.  The merged component
therefore contributes zero, so
$\Delta=-\sum_{K\in\mathcal C}\varphi(K)\le0$.
All six bounds are at most $W/4+1/2$, which
proves~\eqref{eq:ws-boundary-update}.  The constant $1/2$ covers the new
vertex $D$ in the worst case, while every old component other than $T$ is
either unchanged in $Q$ or explicitly subtracted in $\Delta$.

Now condition on the history and on the adaptive rule selecting an
unsatisfied clause $r$.  It is the root of a defect tree $T$.  Put
$W=W_T(r)$ and $d=\deg_T(r)$.  Each of the three literal positions is chosen
with probability $1/3$.

Choosing an edge $ru$ of $T$ moves the defect from $r$ to $u$ without changing
the tree.  The other $3-d$ choices satisfy~\eqref{eq:ws-boundary-update}.
Using~\eqref{eq:ws-rerooting-identity}, we obtain
\begin{align*}
  \E{\Phi_{t+1}\mid\mathcal F_t}
  &\le Q+\frac13\sum_{u\sim_T r}W_T(u)
       +\frac{3-d}{3}\left(\frac14W+\frac12\right)\\
  &=Q+W-\frac{3-d}{12}W-\frac d6.
\end{align*}
Since $\Phi_t=Q+W$, the conditional drift is at most $-1/4$.  Indeed, for
$d=0$ we have $W=1$, and for $d=1,2,3$ the respective bounds are
$-1/3$, $-5/12$, and $-1/2$, using $W\ge1$.

Let $\tau=T_{\WS}$ and set $\widehat\Phi_t=\Phi_{t\wedge\tau}$.  For every
$N\ge1$, telescoping the drift inequality gives
\[
  0\le\E{\widehat\Phi_N}
  \le\E{\Phi_0}-\frac14\sum_{t=0}^{N-1}\Prob(\tau>t)
  =\E{\Phi_0}-\frac14\E{\tau\wedge N}.
\]
By~\eqref{eq:ws-global-potential-bound},
$\E{\tau\wedge N}\le4m$.  Monotone convergence yields
\begin{equation}
  \E{T_{\WS}}\le4m.
  \label{eq:ws-linear-upper-bound}
\end{equation}

For general signs, apply \Cref{prop:ws-splitting} to $F$.
The split formula $F^\sharp$ still has $m$ clauses and satisfies the
opposite-sign case.  Hence~\eqref{eq:ws-linear-upper-bound} gives
\[
  \E{T_{\WS}(F)}\le\E{T_{\WS}(F^\sharp)}\le4m.
\]

The formula has $3m$ occurrences and each of its $n$ variables occurs at most
twice.  Thus $3m\le2n$, which gives $4m\le8n/3$.
\end{proof}

\begin{remark}
The linear dependence on $m$ is tight.  There are formulas with
$\Delta(F)=1$ for which
$T_{\WS}=m$ from a deterministic initial assignment, while under a uniform
initial assignment the same formulas satisfy $\E{T_{\WS}}=m/8$.

Indeed, let
\[
  F_m=\bigwedge_{i=1}^m
    (x_{i,1}\vee x_{i,2}\vee x_{i,3}),
\]
where all variables are distinct.  From the all-zero assignment, each update
permanently satisfies exactly one clause, so $T_{\WS}=m$.  Under a uniform
initial assignment, $T_{\WS}$ is binomial with parameters $m$ and $1/8$.
Therefore $\E{T_{\WS}}=m/8$.
\end{remark}


\section{A Hard Family of Bounded-Occurrence 3-CNF Formulas}
\label{sec:joint-hard-family}

In this section, we prove \Cref{thm-bounded-three}.
We construct a family of 3-CNF formulas $F$ with
$\Delta(F)=3$ in which every clause contains three distinct variables and
both WalkSAT and the MT algorithm have expected running time
$2^{\Theta(\sqrt m)}$.  We also give a
completion in which every variable occurs exactly three times.  
The precise
bounds are as follows, which immediately implies \Cref{thm-bounded-three}.

\begin{theorem}
\label{thm:proper-joint-hard-family}
For every integer $L\ge40$, there are satisfiable 3-CNF formulas $F_L$ and
$\widehat F_L$, each with $\Theta(L^2)$ clauses, maximum variable occurrence
three, and three distinct variables per clause.  For either formula $F$
with $m$ clauses and either algorithm $\mathcal A\in\{\WS,\MT\}$,
\begin{equation}
 2^{\sqrt m/32}
 \le \E{T_{\mathcal A}(F)}
 \le 2^{3\sqrt m}.
 \label{eq:proper-joint-bounds}
\end{equation}
Moreover, $F_L$ has no repeated clauses, and every variable in
$\widehat F_L$ occurs once positively and twice negatively.
\end{theorem}

The lower bound uses the uniform product distribution.  The upper bound
holds from every initial assignment, and both bounds hold under every
adaptive clause-selection rule of Section~\ref{sec:model}.
The exact sizes are given in \eqref{eq:proper-family-size}
and \eqref{eq:proper-completed-size}.

\subsection{A Hard Family of 3-CNF Formulas}
\label{subsec:proper-construction}

\paragraph{Construction of $F_L$.}
Fix an integer $L\ge40$.  Introduce variables $p_{j,i}$ for
$j\in[3]$, $i\in[L]$, and $x_{j,i,\ell}$ for $j\in[3]$,
$i\in[L-1]$, $0\le\ell\le4L+1$.  All these variables are distinct.
For every $j\in[3]$ and $i\in[L-1]$, define
\begin{align}
 A_{j,i}&=(\neg p_{j,i}\vee x_{j,i,0}\vee p_{j,i+1}),
 \label{eq:proper-a-clause}\\
 B_{j,i}&=(\neg p_{j,i}\vee\neg x_{j,i,0}\vee x_{j,i,1}),
 \label{eq:proper-b-clause}\\
 C_{j,i,\ell}&=(x_{j,i,\ell}\vee\neg x_{j,i,\ell-1}
                  \vee\neg x_{j,i,\ell-2}),
 \qquad 2\le\ell\le4L+1.
 \label{eq:proper-c-clause}
\end{align}
Let
\begin{equation}
 R=(p_{1,1}\vee p_{2,1}\vee p_{3,1}).
 \label{eq:proper-root-clause}
\end{equation}
Define
\begin{equation}
 F_L\triangleq R\wedge
 \bigwedge_{j=1}^{3}\bigwedge_{i=1}^{L-1}
 \left(A_{j,i}\wedge B_{j,i}\wedge
       \bigwedge_{\ell=2}^{4L+1}C_{j,i,\ell}\right).
 \label{eq:proper-formula}
\end{equation}
We call $R$ the \emph{root clause}, the variables $p_{j,i}$ the
\emph{connecting variables}, and, for each $(j,i)$, the collection
$A_{j,i},B_{j,i},C_{j,i,2},\ldots,C_{j,i,4L+1}$ a \emph{module},
whose \emph{tail} consists of $x_{j,i,2},\ldots,x_{j,i,4L+1}$.
The index $j$ identifies one of the three branches.

\begin{proposition}
\label{prop:proper-family-properties}
The formula $F_L$ constructed above is satisfiable,
has no repeated clauses, and satisfies $\Delta(F_L)=3$.  Every clause
contains exactly three distinct variables.  Its numbers of clauses and
variables are, respectively,
\begin{equation}
 m=12L^2-6L-5,
 \qquad
 n=12L^2-3L-6.
 \label{eq:proper-family-size}
\end{equation}
\end{proposition}

\begin{proof}
There are $3(L-1)$ modules.  Each has $4L+2$ auxiliary variables and
$4L+2$ clauses: one $A$-clause, one $B$-clause, and $4L$ $C$-clauses.
Including the root clause and the $3L$ connecting variables gives
\[
 m=1+3(L-1)(4L+2)=12L^2-6L-5,
 \qquad n=3L+3(L-1)(4L+2)=12L^2-3L-6.
\]

Every clause contains three distinct variables by construction.
The root clause and clauses of types $A,B,C$ contain respectively
three, two, one, and zero connecting variables, so clauses of different
types cannot coincide.  Within each type, different indices give
different sets of variables.  Hence $F_L$ has no repeated clauses.

Each $p_{j,i}$ with $i<L$ occurs once
positively and twice negatively, whereas $p_{j,L}$ occurs only once,
positively.  In each module, $x_{j,i,0},\ldots,x_{j,i,4L-1}$ also occur
once positively and twice negatively.  The variable $x_{j,i,4L}$ occurs
once with each sign, and $x_{j,i,4L+1}$ occurs only once, positively.
Hence $\Delta(F_L)=3$.

Finally, consider the assignment
\begin{equation}
 p_{1,i}=1,\qquad p_{2,i}=p_{3,i}=0,\qquad x_{j,i,\ell}=0
 \quad\text{for all applicable indices}
 \label{eq:proper-satisfying-assignment}
\end{equation}
The root clause is satisfied by $p_{1,1}=1$.  Each $A$-clause in branch
$1$ is satisfied by $p_{1,i+1}=1$, and each $A$-clause in branches $2,3$
is satisfied by $\neg p_{j,i}$.  Every $B$-clause is satisfied by
$\neg x_{j,i,0}$, and every $C$-clause is satisfied by its negative
literals.  Thus this assignment satisfies $F_L$.
\end{proof}

We next describe the structure of satisfying assignments needed for the
potential argument.

\begin{lemma}[The structure of satisfying assignments]
\label{lem:proper-terminal-structure}
Every satisfying assignment of $F_L$ has at least one of the following
properties:
\begin{enumerate}
 \item For some $j\in[3]$, we have $p_{j,i}=1$ for every $i\in[L]$ and
 $x_{j,i,0}\le x_{j,i,1}$ for every $i\in[L-1]$.
 \item For some $j\in[3]$ and $i\in[L-1]$, we have
 $x_{j,i,\ell}=1$ for every $\ell\in\{0,\ldots,4L+1\}$.
\end{enumerate}
\end{lemma}

\begin{proof}
The root clause guarantees that $p_{j,1}=1$ for some $j\in[3]$.
If $p_{j,i}=1$ for all $i\in[L]$, each clause $B_{j,i}$ forces
$x_{j,i,0}\le x_{j,i,1}$, giving the first property.
Otherwise, there is an index $i\in[L-1]$ with
$p_{j,i}=1$ and $p_{j,i+1}=0$.  The clause $A_{j,i}$ forces
$x_{j,i,0}=1$, and $B_{j,i}$ then forces $x_{j,i,1}=1$.
Induction using $C_{j,i,2},\ldots,C_{j,i,4L+1}$ forces every remaining
auxiliary variable of the module to be $1$, giving the second property.
\end{proof}

\subsection{The potential and its local changes}

We use the same potential for WalkSAT and MT.  We first define the branch
and module factors and establish their ranges and zero conditions.
Lemma~\ref{lem:proper-terminal-structure} will then imply that their
product vanishes on every satisfying assignment.  The local drift bounds
are proved afterwards.

First, for each branch $j\in[3]$ and assignment $X$, define
\begin{equation}
\begin{aligned}
 K_j(X)&\triangleq
 \left(
 \sum_{i=1}^{L}\left(\frac32\right)^{i-1}(1-p_{j,i})
 +\frac12\sum_{i=1}^{L-1}\left(\frac32\right)^{i-1}
       (x_{j,i,0}-x_{j,i,1})
 \right)^+,\qquad j\in[3],\\
 M&\triangleq3\left(\left(\frac32\right)^L-1\right),
\end{aligned}
\label{eq:proper-chain-potential}
\end{equation}
where $z^+=\max\{z,0\}$ and all variables on the right are evaluated at
$X$.

\begin{proposition}[The chain factors]
\label{prop:proper-chain-properties}
For every assignment $X$ and every $j\in[3]$,
\begin{equation}
 0\le K_j(X)\le M.
 \label{eq:proper-chain-range}
\end{equation}
If $p_{j,i}=1$ for every $i\in[L]$ and $x_{j,i,0}\le x_{j,i,1}$
for every $i\in[L-1]$, then $K_j(X)=0$.
\end{proposition}

\begin{proof}
The positive part ensures nonnegativity.  Since $1-p_{j,i}\le1$ and
$x_{j,i,0}-x_{j,i,1}\le1$, we have
\[
 K_j(X)\le
 \sum_{i=1}^{L}\left(\frac32\right)^{i-1}
 +\frac12\sum_{i=1}^{L-1}\left(\frac32\right)^{i-1}
 \le\frac32\sum_{i=1}^{L}\left(\frac32\right)^{i-1}=M.
\]
Under the stated condition, every term $1-p_{j,i}$ is zero and every
term $x_{j,i,0}-x_{j,i,1}$ is nonpositive.  The expression inside the
positive part in \eqref{eq:proper-chain-potential} is therefore
nonpositive, so $K_j(X)=0$.
\end{proof}

Next, fix a module $(j,i)$.  Its tail consists of the $4L$ variables
$x_{j,i,2},\ldots,x_{j,i,4L+1}$.  Partition these variables into four
consecutive blocks of length $L$, so block $b\in[4]$ consists of
\[
 x_{j,i,\,2+(b-1)L},\ldots,x_{j,i,\,1+bL}.
\]
Define the normalized weighted number of zeros in each block by
\begin{align}
 q_{j,i,b}(X)
 &\triangleq\frac{3}{2M}\sum_{u=0}^{L-1}\left(\frac32\right)^u
       (1-x_{j,i,\,2+(b-1)L+u}),
 \label{eq:proper-block-potential}
\end{align}
Combine the four block factors into the module factor
\begin{equation}
 Q_{j,i}(X)\triangleq1-\prod_{b=1}^{4}(1-q_{j,i,b}(X)).
 \label{eq:proper-module-potential}
\end{equation}
\begin{proposition}[The module factors]
\label{prop:proper-module-properties}
For every assignment $X$, every module $(j,i)$, and every $b\in[4]$,
\[
 0\le q_{j,i,b}(X)\le1,
 \qquad 0\le Q_{j,i}(X)\le1.
\]
Moreover,
\begin{equation}
 Q_{j,i}(X)=0
 \quad\Longleftrightarrow\quad
 x_{j,i,2}=\cdots=x_{j,i,4L+1}=1.
 \label{eq:proper-module-zero}
\end{equation}
\end{proposition}

\begin{proof}
The total weight in a block is
$\sum_{u=0}^{L-1}(3/2)^u=2M/3$.  Thus $q_{j,i,b}$ is a weighted average
of bits $1-x_{j,i,\ell}$ and belongs to $[0,1]$.  All weights are
positive, so $q_{j,i,b}(X)=0$ exactly when every variable of block $b$
is $1$.

Each factor $1-q_{j,i,b}(X)$ belongs to $[0,1]$, which also gives
$0\le Q_{j,i}(X)\le1$.  Their product equals $1$ exactly when every
factor equals $1$, or equivalently when all four block potentials are
zero.  Since the blocks partition the tail, this proves
\eqref{eq:proper-module-zero}.
\end{proof}

Finally, define the global potential by
\begin{equation}
 P(X)\triangleq K_1(X)K_2(X)K_3(X)
       \prod_{j=1}^{3}\prod_{i=1}^{L-1}Q_{j,i}(X).
 \label{eq:proper-global-potential}
\end{equation}

\begin{proposition}[The product potential]
\label{prop:proper-product-properties}
The factors in \eqref{eq:proper-global-potential} depend on pairwise
disjoint sets of variables.  For every assignment $X$ and every
satisfying assignment $Y$ of $F_L$,
\[
 0\le P(X)\le M^3,
 \qquad P(Y)=0.
\]
\end{proposition}

\begin{proof}
Each $K_j$ depends only on the connecting variables of branch $j$ and
the first two auxiliary variables of its modules.  Each $Q_{j,i}$ depends
only on the tail variables of its own module.  Thus the factors in $P$
depend on pairwise disjoint sets of variables.
Propositions~\ref{prop:proper-chain-properties}
and~\ref{prop:proper-module-properties} give $0\le P(X)\le M^3$,
since each of the three branch factors is at most $M$ and every module
factor is at most $1$.

If $Y$ satisfies $F_L$, apply Lemma~\ref{lem:proper-terminal-structure}.
In its first case, Proposition~\ref{prop:proper-chain-properties} gives
$K_j(Y)=0$ for the indicated branch.  In its second case,
Proposition~\ref{prop:proper-module-properties} gives $Q_{j,i}(Y)=0$
for the indicated module.  In either case, a factor of $P(Y)$ is zero.
\end{proof}

The next two lemmas give the local calculations used in both lower bounds.
Throughout these lemmas, $X$ is a fixed assignment, $C$ is a fixed
unsatisfied clause, and $X'$ is the assignment after updating $C$.
Subscripts $\WS$ and $\MT$ on expectations specify the update rule.

\begin{lemma}[Changes in the chain potential]
\label{lem:proper-chain-drift}
For either algorithm and every non-root clause $C$ of $F_L$, we have
\[
 \E{K_j(X')-K_j(X)\mid X,C}\ge0
 \qquad\text{for every }j\in[3].
\]
If $R$ is unsatisfied, changing $p_{j,1}$ from $0$ to $1$ decreases
$K_j$ by at most $1$ and leaves the other chain factors unchanged.
\end{lemma}

\begin{proof}
An update in a different branch leaves $K_j$ unchanged.  For a clause in
branch $j$, consider first the affine expression inside the positive part
in \eqref{eq:proper-chain-potential}.

If $A_{j,i}$ is unsatisfied, then
$p_{j,i}=1$ and $x_{j,i,0}=p_{j,i+1}=0$.
Flipping these three variables changes this expression by
$(3/2)^{i-1}$, $\tfrac12(3/2)^{i-1}$, and $-(3/2)^i$, respectively.
Their sum is
\[
 \left(\frac32\right)^{i-1}
 +\frac12\left(\frac32\right)^{i-1}
 -\left(\frac32\right)^i=0.
\]
If $B_{j,i}$ is unsatisfied, then
$p_{j,i}=x_{j,i,0}=1$ and $x_{j,i,1}=0$.
The three changes are $(3/2)^{i-1}$,
$-\tfrac12(3/2)^{i-1}$, and $-\tfrac12(3/2)^{i-1}$, again summing to zero.

If $C_{j,i,2}$ is unsatisfied, then
$x_{j,i,0}=x_{j,i,1}=1$ and $x_{j,i,2}=0$.
Flipping the first two variables changes the affine expression by
$-\tfrac12(3/2)^{i-1}$ and $\tfrac12(3/2)^{i-1}$, while flipping the
third has no effect.  If $C_{j,i,3}$ is unsatisfied, its only variable
appearing in the expression is $x_{j,i,1}=1$; flipping it increases the
expression by $\tfrac12(3/2)^{i-1}$.  Clauses $C_{j,i,\ell}$ with
$\ell\ge4$ do not change the expression.

For WalkSAT, the expected change in this affine expression is one third
of the sum of its three single-variable changes.  For MT, each of the
three variables flips with probability $1/2$, so linearity gives one
half of the same sum.  The expected change is therefore nonnegative
for both algorithms.  Since $z\mapsto z^+$ is convex and nondecreasing,
Jensen's inequality gives the asserted inequality for $K_j$.

Finally, when $R$ is unsatisfied, $p_{1,1}=p_{2,1}=p_{3,1}=0$.
Changing $p_{j,1}$ to $1$ decreases the affine expression for $K_j$ by
exactly $1$.  Taking its positive part decreases the result by a number
in $[0,1]$, proving the root assertion.
\end{proof}

\begin{lemma}[Changes in the module potential]
\label{lem:proper-module-drift}
If $C=C_{j,i,\ell}$ is unsatisfied, then
\begin{align*}
 \E[\WS]{Q_{j,i}(X')-Q_{j,i}(X)\mid X,C}&\ge-\frac1{2M},\\
 \E[\MT]{Q_{j,i}(X')-Q_{j,i}(X)\mid X,C}&\ge-\frac3{4M}.
\end{align*}
All other module potentials are unchanged.  Updating a clause of type
$A,B$, or $R$ leaves every module potential unchanged.
\end{lemma}

\begin{proof}
Fix the module and suppress its indices $j,i$ in this proof.
Since $C_\ell$ is unsatisfied, $x_\ell=0$ and
$x_{\ell-1}=x_{\ell-2}=1$.  Thus flipping $x_\ell$ decreases the
weighted sum defining its block potential, whereas flipping either
predecessor increases the weighted sum of the block containing that
predecessor, if any.

For a block containing $x_\ell$, write
$\ell=2+(b-1)L+u$, where $0\le u\le L-1$.
Before multiplication by $3/(2M)$, the sum of the three single-variable
changes in this block is
\[
 \begin{cases}
 -1, & u=0,\\
 1-\dfrac32=-\dfrac12, & u=1,\\
 -\left(\dfrac32\right)^u
 +\left(\dfrac32\right)^{u-1}
 +\left(\dfrac32\right)^{u-2}
 =\dfrac14\left(\dfrac32\right)^{u-2}, & u\ge2.
 \end{cases}
\]
In every other block the sum is nonnegative.  Consequently, at most one
block has a negative expected change.  Its magnitude is at most
$1/(2M)$ for WalkSAT and $3/(4M)$ for MT, since their expected changes
in an affine expression are respectively one third and one half of the
sum of the three single-variable changes.

For WalkSAT, a single update changes at most one block.  Because $Q$ is
affine in each block potential when the others are fixed,
\[
 \E[\WS]{Q(X')-Q(X)\mid X,C_\ell}
 =\sum_{b=1}^{4}\left(\prod_{c\ne b}(1-q_c(X))\right)
       \E[\WS]{q_b(X')-q_b(X)\mid X,C_\ell}.
\]
Every coefficient belongs to $[0,1]$, so the sum is at least $-1/(2M)$.

For MT, conditional on $X$ and $C_\ell$, different blocks depend on
disjoint sets of fresh independent bits.  Their updated potentials are
therefore conditionally independent, and
\[
 \E[\MT]{Q(X')\mid X,C_\ell}
 =1-\prod_{b=1}^{4}\left(1-\E[\MT]{q_b(X')\mid X,C_\ell}\right).
\]
On $[0,1]^4$, the expression $1-\prod_b(1-q_b)$ is nondecreasing in
each coordinate, with slope at most $1$ in that coordinate.  At most one
coordinate decreases when the old block potentials are replaced by their
conditional expectations, and it decreases by at most $3/(4M)$.
This proves the MT inequality.  The remaining assertions follow from
the disjoint variable sets used by the module potentials.
\end{proof}

\subsection{The expected time of WalkSAT}

In this subsection, we prove the following lower bound.

\begin{proposition}
\label{prop:proper-walksat-lower}
Starting from the uniform product distribution,
\begin{equation}
 \E{T_{\WS}(F_L)}
 \ge\frac{(3/2)^L-1}{9}
       \left(\frac{15}{16}\right)^{3(L-1)}.
 \label{eq:proper-walksat-lower}
\end{equation}
\end{proposition}

\begin{proof}
Let $X_t$ denote the assignment after $t$ updates, and let $\mathcal F_t$
denote the history through $X_t$.  We use the potential $P$ in
\eqref{eq:proper-global-potential}.  Recall that $0\le P(X)\le M^3$
for every assignment $X$ and $P(Y)=0$ for every satisfying assignment $Y$.
We first bound the conditional expected loss in $P$ for each possible
type of selected clause.

If $A_{j,i}$ or $B_{j,i}$ is selected, only the factor $K_j$ can change.
All other factors in $P$ are fixed and nonnegative.  Hence
Lemma~\ref{lem:proper-chain-drift} gives
\[
 \E{P(X_{t+1})-P(X_t)\mid\mathcal F_t,\
      A_{j,i}\text{ or }B_{j,i}\text{ selected}}\ge0.
\]
If $C_{j,i,\ell}$ is selected, only $K_j$ and $Q_{j,i}$ can change.
Their variable sets are disjoint, so a single flip changes at most one
of these two factors.  Expanding the product using the old values of the
other factors, the contribution from $K_j$ has nonnegative conditional
expectation by Lemma~\ref{lem:proper-chain-drift}.  The possible loss
from $Q_{j,i}$ is bounded by Lemma~\ref{lem:proper-module-drift}, giving
\begin{equation}
\begin{aligned}
 &\E{P(X_t)-P(X_{t+1})\mid\mathcal F_t,\ C_{j,i,\ell}\text{ selected}}\\
 &\qquad\le\frac{K_1(X_t)K_2(X_t)K_3(X_t)}{2M}
       \prod_{(a,h)\ne(j,i)}Q_{a,h}(X_t)
 \le\frac{M^2}{2}.
\end{aligned}
\label{eq:proper-walksat-tail-loss}
\end{equation}
The product here ranges over all modules other than $(j,i)$.

If $R$ is selected, all three root variables are $0$.  WalkSAT flips
one of them, say $p_{j,1}$, decreasing $K_j$ by at most $1$ by
Lemma~\ref{lem:proper-chain-drift}.  All other factors remain unchanged.
Consequently, for every possible outcome of this update,
\[
 P(X_t)-P(X_{t+1})
 \le\prod_{a\ne j}K_a(X_t)
       \prod_{a=1}^{3}\prod_{h=1}^{L-1}Q_{a,h}(X_t)
 \le M^2.
\]
Combining the three cases, for every $t<T_{\WS}(F_L)$ we obtain
\begin{equation}
 \E{P(X_t)-P(X_{t+1})\mid\mathcal F_t}\le M^2.
 \label{eq:proper-walksat-step-loss}
\end{equation}
This bound holds for any adaptive choice of an unsatisfied clause,
because it holds after conditioning on each such choice.

It remains to estimate the initial potential.  Each variable is initially
an independent uniform bit.  Since $z^+\ge z$, definition
\eqref{eq:proper-chain-potential} gives
\begin{equation}
 \E{K_j(X_0)}
 \ge\frac12\sum_{i=1}^{L}\left(\frac32\right)^{i-1}
 =\frac M3.
 \label{eq:proper-chain-initial}
\end{equation}
Indeed, every term $x_{j,i,0}-x_{j,i,1}$ has expectation zero.
Likewise, \eqref{eq:proper-block-potential} gives
$\E{q_{j,i,b}(X_0)}=1/2$.  The four blocks of a module have disjoint
variable sets and are initially independent, so
\begin{equation}
 \E{Q_{j,i}(X_0)}
 =1-\prod_{b=1}^{4}\left(1-\E{q_{j,i,b}(X_0)}\right)
 =\frac{15}{16}.
 \label{eq:proper-module-initial}
\end{equation}
Finally, all chain factors and all module factors depend on pairwise
disjoint sets of initial bits.  Therefore,
\begin{equation}
\begin{aligned}
 \E{P(X_0)}
 &=\prod_{j=1}^{3}\E{K_j(X_0)}
       \prod_{j=1}^{3}\prod_{i=1}^{L-1}\E{Q_{j,i}(X_0)}\\
 &\ge\frac{M^3}{27}\left(\frac{15}{16}\right)^{3(L-1)}.
\end{aligned}
\label{eq:proper-initial-product}
\end{equation}

If $\E{T_{\WS}(F_L)}=\infty$, the proposition is immediate.
Otherwise the hitting time is integrable, and
Lemma~\ref{lem:drift-lower-bound}, together with
\eqref{eq:proper-walksat-step-loss} and \eqref{eq:proper-initial-product},
gives
\[
 \E{T_{\WS}(F_L)}
 \ge\frac{\E{P(X_0)}}{M^2}
 \ge\frac M{27}\left(\frac{15}{16}\right)^{3(L-1)}.
\]
Substituting the value of $M$ proves \eqref{eq:proper-walksat-lower}.
\end{proof}

\subsection{The expected time of MT}

For MT we use the same potential and the same initial estimate.  The
additional point is to account for updates that change several factors
at once.

\begin{proposition}
\label{prop:proper-mt-lower}
Starting from the uniform product distribution,
\begin{equation}
 \E{T_{\MT}(F_L)}
 \ge\frac{2((3/2)^L-1)}{27}
       \left(\frac{15}{16}\right)^{3(L-1)}.
 \label{eq:proper-mt-lower}
\end{equation}
\end{proposition}

\begin{proof}
Let $X_t$ and $\mathcal F_t$ denote the assignment and history for MT.
Every resampling counts as one update, including a resampling that leaves
the assignment unchanged.  As in the WalkSAT proof, updating an
unsatisfied $A_{j,i}$ or $B_{j,i}$ has nonnegative conditional expected
change in $P$, since only $K_j$ can change.

Suppose that $C=C_{j,i,\ell}$ is selected.  The only factors that may
change are $K_j$ and $Q_{j,i}$.  Conditional on $\mathcal F_t$ and this
choice of $C$, they depend on disjoint sets of fresh resampling bits.
They are therefore conditionally independent after the update.
For brevity, in the following calculation all expectations are conditional
on $\mathcal F_t$ and the selection of $C$, and we write $X=X_t$ and
$X'=X_{t+1}$.  Then
\begin{align*}
 &\E{K_j(X')Q_{j,i}(X')}-K_j(X)Q_{j,i}(X)\\
 &\quad=\bigl(\E{K_j(X')}-K_j(X)\bigr)Q_{j,i}(X)
       +\E{K_j(X')}\bigl(\E{Q_{j,i}(X')}-Q_{j,i}(X)\bigr)\\
 &\quad\ge-\frac34.
\end{align*}
The last inequality uses Lemmas~\ref{lem:proper-chain-drift}
and~\ref{lem:proper-module-drift}, together with $0\le K_j\le M$.
The remaining two chain factors are at most $M$, and every remaining
module factor is at most $1$.  Hence
\begin{equation}
 \E{P(X_t)-P(X_{t+1})\mid\mathcal F_t,\ C_{j,i,\ell}\text{ selected}}
 \le\frac{3M^2}{4}.
 \label{eq:proper-mt-tail-loss}
\end{equation}

If $R$ is selected, its three variables are initially $0$ and are
resampled to independent uniform bits.  For every outcome,
\[
 0\le K_j(X_t)-K_j(X_{t+1})\le X_{t+1}(p_{j,1}),
 \qquad j\in[3].
\]
The module potentials remain unchanged.  Expanding the loss in the
product of the three chain factors one factor at a time, and using
$0\le K_j\le M$, gives
\[
 P(X_t)-P(X_{t+1})
 \le M^2\sum_{j=1}^{3}X_{t+1}(p_{j,1}).
\]
Each bit on the right has conditional expectation $1/2$.  Thus
\[
 \E{P(X_t)-P(X_{t+1})\mid\mathcal F_t,\ R\text{ selected}}
 \le\frac{3M^2}{2}.
\]
Combining all clause types yields, before termination,
\begin{equation}
 \E{P(X_t)-P(X_{t+1})\mid\mathcal F_t}\le\frac{3M^2}{2}.
 \label{eq:proper-mt-step-loss}
\end{equation}

The initial distribution is the same as for WalkSAT, so
\eqref{eq:proper-initial-product} remains valid.  An infinite expected
hitting time already implies the desired bound.  In the finite case,
Lemma~\ref{lem:drift-lower-bound} applies and gives
\[
 \E{T_{\MT}(F_L)}
 \ge\frac{2\E{P(X_0)}}{3M^2}
 \ge\frac{2M}{81}\left(\frac{15}{16}\right)^{3(L-1)}.
\]
Substitution of $M$ proves \eqref{eq:proper-mt-lower}.
\end{proof}

\subsection{Completing the occurrence counts}

We now construct $\widehat F_L$ by adding clauses to $F_L$ without
introducing new variables.  Enumerate the $3(L-1)$ modules in any fixed
order.  For module $\ell$, denote its last two variables by $u_\ell$
and $v_\ell$, respectively, and use cyclic indexing
$v_{3(L-1)+1}=v_1$.  Add the clauses
\begin{align}
 D_\ell&=(\neg u_\ell\vee\neg v_\ell\vee\neg v_{\ell+1}),
 \qquad \ell\in[3(L-1)],
 \label{eq:proper-completion-tail}\\
 E_1=E_2&=(\neg p_{1,L}\vee\neg p_{2,L}\vee\neg p_{3,L}).
 \label{eq:proper-completion-end}
\end{align}
The clauses $E_1,E_2$ have different indices, even though their literals
are identical.  Each $D_\ell$ contains three distinct variables, since
there is more than one module.

Every $u_\ell$ acquires one negative occurrence, every $v_\ell$ acquires
two negative occurrences, and every $p_{j,L}$ acquires two negative
occurrences.  The occurrence counts of all other variables are unchanged.
Thus every variable in $\widehat F_L$ occurs once positively and twice
negatively.  We have added $3(L-1)+2=3L-1$ clauses, so
\begin{equation}
 |\widehat F_L|=12L^2-3L-6=n.
 \label{eq:proper-completed-size}
\end{equation}
The assignment in \eqref{eq:proper-satisfying-assignment} satisfies every
new clause as well: all tail variables are $0$, and $p_{2,L}=p_{3,L}=0$.

\begin{proposition}
\label{prop:proper-completion-lower}
The lower bounds in Propositions~\ref{prop:proper-walksat-lower}
and~\ref{prop:proper-mt-lower} also hold with $F_L$ replaced by
$\widehat F_L$.
\end{proposition}

\begin{proof}
Use the same potential $P$ on the same set of variables.  Every satisfying
assignment of $\widehat F_L$ satisfies $F_L$, so its potential is zero.
The uniform product distribution at initialization and hence
\eqref{eq:proper-initial-product} are unchanged.

It remains to check updates of the added clauses.  An unsatisfied
$D_\ell$ has all three of its variables equal to $1$.  Under either
algorithm, updating it can only change some of these variables from $1$
to $0$.  Such changes increase the corresponding block potentials and
cannot decrease any module potential.  All chain factors remain fixed.
Similarly, an unsatisfied $E_1$ or $E_2$ has
$p_{1,L}=p_{2,L}=p_{3,L}=1$.  Updating it can only change some of these
bits to $0$, which cannot decrease any $K_j$ and leaves every module
factor unchanged.  Therefore every update of an added clause satisfies
$P(X_{t+1})\ge P(X_t)$ for each individual outcome.

The expected-loss bounds \eqref{eq:proper-walksat-step-loss} and
\eqref{eq:proper-mt-step-loss} thus remain valid after completion.
The same drift arguments prove both lower bounds.
\end{proof}

We can now express the lower bounds in terms of the full formula size.
For either formula $F$ and either algorithm $\mathcal A$, the weaker of
the two estimates gives
\begin{equation}
 \E{T_{\mathcal A}(F)}
 \ge\frac{2((3/2)^L-1)}{27}
       \left(\frac{15}{16}\right)^{3(L-1)}.
 \label{eq:proper-common-lower}
\end{equation}
For $L\ge2$, $(3/2)^L-1\ge(3/2)^L/2$, and
\[
 \frac32\left(\frac{15}{16}\right)^3
 =\frac{10125}{8192}>\frac65>2^{1/4}.
\]
Consequently, the right-hand side of \eqref{eq:proper-common-lower}
is at least
\[
 \frac1{27}
 \left(\frac32\left(\frac{15}{16}\right)^3\right)^L
 >\frac{2^{L/4}}{27}\ge2^{L/8},
\]
where the last inequality uses $L\ge40$.
Both formulas have $m<16L^2$ clauses, so
\begin{equation}
 \E{T_{\mathcal A}(F)}\ge2^{L/8}\ge2^{\sqrt m/32}.
 \label{eq:proper-size-lower}
\end{equation}

\subsection{A matching upper bound}

We finish by proving an upper bound that holds for every initial assignment
and every adaptive clause-selection rule, for both formulas and both
algorithms.  Write $\E[\sigma]{\cdot}$ for expectation from the fixed initial
assignment $\sigma$.

\begin{proposition}
\label{prop:proper-joint-upper}
For $F\in\{F_L,\widehat F_L\}$ with $n$ variables, every initial
assignment $\sigma$, and $\mathcal A\in\{\WS,\MT\}$, we have
\begin{equation}
 \E[\sigma]{T_{\mathcal A}(F)}
 \le3n+75\left(5^{3L-1}-1\right).
 \label{eq:proper-parameter-upper}
\end{equation}
\end{proposition}

The proof uses two partial assignments that satisfy every clause involving
one of their assigned variables.  Such a partial assignment is called an
\emph{autarky}, and its support is the set of variables it assigns.
For $j\in\{1,3\}$, define the partial assignment $\alpha_j$ by
\begin{equation}
 \alpha_j:\quad
 p_{j,i}=1,\quad p_{2,i}=0\quad(i\in[L]),
 \qquad x_{j,i,0}=0\quad(i\in[L-1]).
 \label{eq:proper-autarkies}
\end{equation}
Each support has size $3L-1$.

To verify the autarky property, consider first branch $j$.  The root and
all its $A$-clauses are satisfied by the assigned connecting variables.
Its $B$-clauses and $C_{j,i,2}$ are satisfied by $\neg x_{j,i,0}$.
On branch $2$, the $A$- and $B$-clauses are satisfied by $\neg p_{2,i}$.
These are all the original clauses touching the support.  For the
completion, no $D_\ell$ touches the support, and both $E_1,E_2$ are
satisfied by $\neg p_{2,L}$.  Thus $\alpha_1$ and $\alpha_3$ are
autarkies of both formulas.

Let $D_j(X)$ be the number of variables in the support of $\alpha_j$
on which $X$ disagrees with $\alpha_j$.  Define
\begin{equation}
 g(z)\triangleq\frac{25}{2}\left(5^{3L-1}-5^{3L-1-z}\right),
 \qquad 0\le z\le3L-1.
 \label{eq:proper-distance-transform}
\end{equation}
The following calculation is valid for either of the two autarkies.

\begin{lemma}
\label{lem:proper-autarky-drift}
Fix $j\in\{1,3\}$, a current assignment $X$, and an unsatisfied clause
$C$.  If $C$ does not touch the support of $\alpha_j$, updating $C$
leaves $D_j$ unchanged.  Otherwise, for either algorithm,
\begin{equation}
 \E{g(D_j(X'))-g(D_j(X))\mid X,C}\le-1.
 \label{eq:proper-autarky-decrease}
\end{equation}
\end{lemma}

\begin{proof}
The first assertion is immediate.  For the second, the autarky property
guarantees a literal of $C$ that is satisfied by $\alpha_j$.  Since $C$
is currently unsatisfied, the variable of that literal disagrees with
$\alpha_j$ under $X$.  In particular, $z=D_j(X)\ge1$.

For WalkSAT, flipping this variable decreases the distance by $1$;
flipping either of the other two variables increases it by at most $1$.
The function $g$ is increasing, and
\[
 g(z)-g(z-1)=50\cdot5^{3L-1-z}.
\]
If $z<3L-1$, the conditional expected change is consequently at most
\[
 \frac{-50+2\cdot10}{3}\,5^{3L-1-z}
 =-10\cdot5^{3L-1-z}\le-1.
\]
If $z=3L-1$, no flip can increase the distance, and the expected change
is at most $-50/3$.

For MT, let $a\le3$ be the number of variables of $C$ in the support,
and let $b\ge1$ be the number of those variables currently disagreeing
with $\alpha_j$.  After resampling, each of the $a$ variables disagrees
with probability $1/2$, independently.  Conditional on $X$ and $C$,
\[
 D_j(X')\overset{d}=z-b+\operatorname{Bin}(a,1/2).
\]
It follows that
\begin{align*}
 \E{g(D_j(X'))-g(D_j(X))\mid X,C}
 &=\frac{25}{2}\,5^{3L-1-z}
       \left(1-5^b\left(\frac35\right)^a\right)\\
 &\le\frac{25}{2}\,5^{3L-1-z}
       \left(1-\frac{27}{25}\right)\\
 &=-5^{3L-1-z}\le-1.
\end{align*}
This proves the assertion for MT as well.
\end{proof}

\begin{proof}[Proof of Proposition~\ref{prop:proper-joint-upper}]
Define the nonnegative potential
\begin{equation}
 \mathcal U(X)\triangleq3\sum_v X(v)+3g(D_1(X))+3g(D_3(X)),
 \label{eq:proper-upper-potential}
\end{equation}
where the sum ranges over all $n$ variables of $F$.
By Lemma~\ref{lem:proper-autarky-drift}, each distance term has
nonpositive conditional drift for every selected clause, and decreases
in conditional expectation by at least $3$ whenever the clause touches
the corresponding support.  We check that these terms compensate for
any increase in the first term.

If $R$ is selected, its three variables are $0$.  The conditional
expected increase in $3\sum_v X(v)$ is $3$ for WalkSAT and $9/2$ for
MT.  The root touches both autarkies, so the conditional expected change
in $\mathcal U$ is at most $3-6=-3$ for WalkSAT and $9/2-6=-3/2$ for MT.

If an $A$-clause is selected, its two positive literals have variables
equal to $0$ and its negative literal has variable equal to $1$.
The conditional expected increase in $3\sum_v X(v)$ is $1$ for
WalkSAT and $3/2$ for MT.  Every $A$-clause touches at least one
autarky: those in branch $1$ touch $\alpha_1$, those in branch $3$
touch $\alpha_3$, and those in branch $2$ touch both.
The conditional expected change in $\mathcal U$ is therefore at most
$1-3=-2$ for WalkSAT and $3/2-3=-3/2$ for MT.

Every $B$- or $C$-clause has one positive and two negative literals.
When such a clause is unsatisfied, its variables consist of one $0$ and
two $1$'s.  Updating it changes $3\sum_v X(v)$ in conditional
expectation by $-1$ for WalkSAT and $-3/2$ for MT.  The distance terms
cannot increase in expectation.  Finally, each added $D$- or $E$-clause
has three negative literals.  When it is unsatisfied, all three
variables are $1$, and the first term decreases in expectation by $3$
for WalkSAT and $9/2$ for MT.  Again the distance terms have
nonpositive conditional drift.

Thus, for both algorithms and every $t<T_{\mathcal A}(F)$,
\[
 \E{\mathcal U(X_{t+1})-\mathcal U(X_t)\mid\mathcal F_t}\le-1.
\]
Moreover, $0\le D_j(X)\le3L-1$, so
\[
 0\le\mathcal U(X)\le3n+6g(3L-1)
 =3n+75\left(5^{3L-1}-1\right).
\]
Applying Lemma~\ref{lem:unit-drift} from the fixed initial assignment
$\sigma$ proves \eqref{eq:proper-parameter-upper}.  In particular, both
algorithms terminate almost surely with finite expectation under every
allowed clause-selection rule.
\end{proof}

For $L\ge40$, both formulas satisfy $11L^2\le m<16L^2$ and $n<16L^2$.
Hence Proposition~\ref{prop:proper-joint-upper} gives
\begin{align*}
 \sup_\sigma\E[\sigma]{T_{\mathcal A}(F)}
 &\le3n+75\left(5^{3L-1}-1\right)\\
 &<48L^2+15\cdot125^L
 \le16\cdot125^L
 <2^{7L+4}\le2^{8L}\le2^{3\sqrt m}.
\end{align*}
Here $48L^2\le125^L$ for $L\ge40$, and
$3\sqrt m\ge3\sqrt{11}\,L>8L$.
Combining this estimate with \eqref{eq:proper-size-lower} proves
Theorem~\ref{thm:proper-joint-hard-family}.

\section{A 3-CNF Family with Polynomial WalkSAT and Exponential MT Time}
\label{sec:mt-slow}

In this section we prove \Cref{thm-mt-exponential}.  The construction is the
Tseitin formula on a connected simple 3-regular graph with charge function
identically zero.  For an assignment, let $D$ be the set of graph vertices at
which the corresponding parity constraint is violated.  Under WalkSAT, the
vertices in $D$ evolve as particles in an annihilating random walk.  Under MT,
the continuous-time version of the process $D_t$ is dual, up to a factor-two
time change, to the threshold voter model with threshold~$1$.  We then choose
locally sparse 3-regular graphs on which the associated threshold contact process
survives for exponential time and transfer this lower bound back to MT.

\subsection{Tseitin encoding and the WalkSAT dynamics}
\label{subsec:tseitin-encoding}

Let $G=(U,E)$ be a finite connected simple 3-regular graph, and write $q=|U|$.
Then $|E|=3q/2$.  Introduce one Boolean variable $x_e$ for every edge
$e\in E$.  If the three edges incident to $v\in U$ are $e_1,e_2,e_3$, include
the four clauses
\begin{equation}
\begin{split}
 & (\neg x_{e_1}\vee x_{e_2}\vee x_{e_3}),
 \qquad (x_{e_1}\vee\neg x_{e_2}\vee x_{e_3}),\\
 & (x_{e_1}\vee x_{e_2}\vee\neg x_{e_3}),
 \qquad (\neg x_{e_1}\vee\neg x_{e_2}\vee\neg x_{e_3}).
\end{split}
\label{eq:tseitin-parity-clauses}
\end{equation}
Let $F(G)$ be the conjunction of these clauses over all $v\in U$.  The four
clauses in \eqref{eq:tseitin-parity-clauses} are the CNF encoding of
\[
  x_{e_1}\oplus x_{e_2}\oplus x_{e_3}=0,
\]
so $F(G)$ is the Tseitin formula on $G$ with zero charge at every vertex.

Identify an assignment to the variables $\{x_e:e\in E\}$ with its edge-label
vector $\sigma\in\bits^E$.  Define the set of violated vertices by
\begin{equation}
  D(\sigma)=\left\{v\in U:\bigoplus_{e\ni v}\sigma(e)=1\right\}.
  \label{eq:tseitin-defect-set}
\end{equation}

\begin{lemma}[Structure of the Tseitin formula]
\label{lem:tseitin-structure}
The formula $F(G)$ is satisfiable, every clause contains three distinct
variables, it has $3q/2$ variables and $4q$ clauses, and
\[
  \Delta(F(G))=8.
\]
For every assignment $\sigma\in\bits^E$, a vertex $v$ lies in $D(\sigma)$ if
and only if exactly one of the four clauses associated with $v$ is unsatisfied.
Consequently,
\begin{equation}
  |\Bad_{F(G)}(\sigma)|=|D(\sigma)|,
  \label{eq:tseitin-bad-count}
\end{equation}
and $|D(\sigma)|$ is even.
\end{lemma}

\begin{proof}
The graph has $3q/2$ edges and each of its $q$ vertices contributes four
clauses.  Since $G$ is simple, the three edges incident to a vertex are
distinct.  Each edge has two endpoints, and its variable occurs in all four
local clauses at each endpoint; hence every variable occurs exactly eight
times.  The all-zero assignment satisfies every parity constraint.

At a fixed vertex, the four clauses in
\eqref{eq:tseitin-parity-clauses} are unsatisfied respectively under the four odd
local assignments $100,010,001,111$, and every even local assignment
satisfies all four clauses.  This proves \eqref{eq:tseitin-bad-count}.
Finally, over $\Ftwo$,
\[
  |D(\sigma)|
  \equiv \sum_{v\in U}\ \sum_{e\ni v}\sigma(e)
  =2\sum_{e\in E}\sigma(e)
  =0\pmod 2,
\]
because every edge is counted at both endpoints.
\end{proof}

With this identification, write $D_t=D(X_t)$ for the violated-vertex set
induced by the current SAT assignment.  Suppose that WalkSAT selects the unique
unsatisfied clause associated
with $v\in D_t$ and flips the edge variable $x_e$ for
$e=\{u,v\}$.  Only the parity constraints at $u$ and $v$ change, so
\begin{equation}
  D_{t+1}=D_t\mathbin\triangle\{u,v\}.
  \label{eq:ws-defect-update}
\end{equation}
If $u\notin D_t$, the particle at $v$ moves to $u$; if $u\in D_t$, the two
particles annihilate.  By \Cref{lem:tseitin-structure}, under the uniform
clause-selection rule the selected vertex is uniform in $D_t$, and the chosen
literal occurrence corresponds to a uniformly chosen neighbor of that vertex.
Thus, the induced process is an annihilating random walk on \(G\).

\begin{corollary}[WalkSAT on the Tseitin family]
\label{cor:tseitin-ws-bound}
For every finite connected simple 3-regular graph $G$ on $q$ vertices,
WalkSAT on $F(G)$ satisfies
\[
  \E{T_{\WS}(F(G))}=O(q^4)
\]
from every initial assignment.
\end{corollary}

\begin{proof}
By \Cref{lem:tseitin-structure}, $F(G)$ is a satisfiable Tseitin formula on
$G$.  For satisfiable Tseitin formulas on \(q\)-vertex graphs, WalkSAT has expected running time \(O(q^4)\) \cite[Theorem~5]{jiang2026separation}. The clause- and variable-selection rules agree with those in Section~2, so the bound applies directly to \(F(G)\).
\end{proof}

\subsection{MT dynamics and threshold-voter duality}

Suppose that MT selects the unique unsatisfied clause associated with
$v\in D_t$ and resamples the three incident edge variables.  Conditional on
the current assignment, each of these three variables changes value
independently with probability $1/2$.  Let $S\subseteq N_G(v)$ be the set of
neighbors joined to $v$ by edges whose values change.  Then $S$ is uniform
over the eight subsets of $N_G(v)$.  Toggling these edge values changes the
parity at every vertex of $S$, and also at $v$ exactly when $|S|$ is odd.
Define
\begin{equation}
  \Gamma(v,S)=
  \begin{cases}
    S, & |S|\text{ is even},\\
    S\cup\{v\}, & |S|\text{ is odd}.
  \end{cases}
  \label{eq:mt-gamma}
\end{equation}
One MT step therefore induces
\begin{equation}
  D_{t+1}=D_t\mathbin\triangle\Gamma(v,S).
  \label{eq:mt-defect-update}
\end{equation}

For the lower bound it is convenient to use a continuous-time version of this
chain.  At a state $D$, give each $v\in D$ an independent rate-one clock.
When the clock at $v$ rings, choose $S\subseteq N_G(v)$ uniformly and apply
\eqref{eq:mt-defect-update}.  Let
\[
  \tau_\varnothing=\inf\{t\ge0:D_t=\varnothing\}.
\]

\begin{lemma}[Continuous-time embedding]
\label{lem:mt-discrete-continuous}
The sequence of states observed at successive clock rings has the same law as
the violated-vertex chain induced by the discrete MT algorithm.  Conditional
on a discrete path
\[
  D_0,D_1,\ldots,D_{T_{\MT}}=\varnothing,
\]
the waiting time before step $r+1$ is exponential with rate $|D_r|$.  Hence
\begin{equation}
  \E{\tau_\varnothing}
  =\E{\sum_{r<T_{\MT}}\frac{1}{|D_r|}}
  \le \frac12\E{T_{\MT}}.
  \label{eq:mt-time-comparison}
\end{equation}
\end{lemma}

\begin{proof}
At a nonempty state $D$, the superposed clocks ring at total rate $|D|$ and,
conditional on a ring, the ringing vertex is uniform in $D$.  Together with
the uniform choice of $S$, this is exactly the discrete MT transition rule.
The choice $S=\varnothing$ gives a self-loop in both descriptions.

Conditional on the discrete state sequence, the waiting times between clock
rings are independent exponentials with the stated rates.  Taking conditional
expectations and then using monotone convergence gives the identity in
\eqref{eq:mt-time-comparison}.  By \Cref{lem:tseitin-structure}, every reachable
nonempty set $D_r$ has even cardinality and hence $|D_r|\ge2$.
\end{proof}

Let $\eta_t\in\bits^U$ be the threshold voter model with threshold~$1$:
a vertex flips at rate one if it has at least one neighbor in the opposite
state, and otherwise does not flip.  Let
\[
  \tau_{\mathrm{cons}}
  =\inf\{t\ge0:\eta_t\in\{\boldsymbol0,\boldsymbol1\}\}
\]
be its consensus time.  We establish a duality between the MT defect process and the threshold-one voter model. Our analysis builds on the classical duality framework \cite{cox1991nonlinear,liggett1999stochastic}.

\begin{lemma}[Duality with the threshold voter model]
\label{lem:mt-fourier-duality}
For $\eta\in\bits^U$ and an even set $D\subseteq U$, define
\[
  H(\eta,D)=(-1)^{\sum_{v\in D}\eta(v)}.
\]
Let $L_D$ and $L_\eta$ be the generators of the continuous-time process $D_t$
and the threshold voter model, respectively.  Then
\begin{equation}
  L_DH(\eta,D)=\frac12L_\eta H(\eta,D).
  \label{eq:generator-duality}
\end{equation}
Consequently, for every $t\ge0$,
\begin{equation}
  \E[D]{H(\eta,D_t)}
  =\E[\eta]{H(\eta_{t/2},D)}.
  \label{eq:mt-voter-duality}
\end{equation}
\end{lemma}

\begin{proof}
Fix $v\in D$.  From \eqref{eq:mt-gamma},
\[
  \frac{H(\eta,D\mathbin\triangle\Gamma(v,S))}{H(\eta,D)}
  =(-1)^{\sum_{u\in S}(\eta(u)+\eta(v))}.
\]
Averaging over the uniform subset $S\subseteq N_G(v)$ gives
\begin{align*}
  2^{-3}\sum_{S\subseteq N_G(v)}
  (-1)^{\sum_{u\in S}(\eta(u)+\eta(v))}
  &=\prod_{u\sim v}\frac{1+(-1)^{\eta(u)+\eta(v)}}{2}\\
  &=\one\{\eta(u)=\eta(v)\text{ for every }u\sim v\}.
\end{align*}
Thus the contribution of $v$ to $L_DH$ is $-H(\eta,D)$ exactly when $v$ has a
neighbor in the opposite state, and is zero otherwise.

For the voter model, flipping $\eta(v)$ changes $H(\eta,D)$ only when
$v\in D$; in that case the increment is $-2H(\eta,D)$.  Since the flip rate is
one whenever an opposite-state neighbor exists, the contribution to
$L_\eta H$ is $-2H(\eta,D)$.  Summing over $v\in D$ proves
\eqref{eq:generator-duality}.  Since both state spaces are finite, the
generator identity implies the semigroup relation
\eqref{eq:mt-voter-duality}.
\end{proof}

\begin{lemma}[Extinction time versus consensus time]
\label{lem:mt-voter-times}
Suppose that the edge variables of $F(G)$ are initially independent unbiased
bits, and independently initialize the voter model from the product
$\operatorname{Bernoulli}(1/2)$ distribution.  Then
\begin{equation}
  \tau_\varnothing\ \overset{d}{=}\ 2\tau_{\mathrm{cons}}.
  \label{eq:extinction-consensus-law}
\end{equation}
\end{lemma}

\begin{proof}
View an edge assignment as a vector in $\Ftwo^E$, and let
$\partial:\Ftwo^E\to\Ftwo^U$ be the vertex--edge incidence map.  Every column
of $\partial$ has even weight.  Conversely, if $u$ and $v$ are joined by a
path, the sum of the incidence columns along that path is
$\mathbf e_u+\mathbf e_v$.  Since $G$ is connected, the image of $\partial$
is exactly the even subspace
\[
  \mathcal E=\{D\subseteq U:|D|\text{ is even}\}.
\]
All fibers of a linear map have the same cardinality, so independent uniform
edge variables induce the uniform distribution on $\mathcal E$.

For a uniform voter configuration $\eta_0$ and fixed $D$,
\[
  \E{H(\eta_0,D)}=\one\{D=\varnothing\}.
\]
The orthogonal complement of $\mathcal E$ is
$\{\boldsymbol0,\boldsymbol1\}$: orthogonality to every incidence column forces
equal values across every edge, and $G$ is connected.  Hence, for fixed
$\eta$,
\[
  \E[D_0\sim\Unif(\mathcal E)]{H(\eta,D_0)}
  =\one\{\eta\in\{\boldsymbol0,\boldsymbol1\}\}.
\]
Average \eqref{eq:mt-voter-duality} over independent
$\eta\sim\Unif(\bits^U)$ and $D\sim\Unif(\mathcal E)$.  For every $t\ge0$,
\[
  \Prob(D_t=\varnothing)
  =\Prob(\eta_{t/2}\in\{\boldsymbol0,\boldsymbol1\}).
\]
Both target sets are absorbing, so the two sides are
$\Prob(\tau_\varnothing\le t)$ and
$\Prob(\tau_{\mathrm{cons}}\le t/2)$, respectively.  This proves
\eqref{eq:extinction-consensus-law}.
\end{proof}

\subsection{Exponential survival on locally sparse 3-regular graphs}

For $A\subseteq U$, let $e_G(A)$ be the number of edges of the induced
subgraph $G[A]$.  We need 3-regular graphs satisfying the following local sparsity
condition:
\begin{equation}
  e_G(A)\le\frac{17}{16}|A|
  \qquad
  \text{whenever }|A|\le\beta q.
  \label{eq:local-sparsity}
\end{equation}

\begin{lemma}[A locally sparse 3-regular graph]
\label{lem:locally-sparse-cubic}
There is an absolute constant $\beta>0$ such that, for every sufficiently
large even integer $q$, there exists a connected simple 3-regular graph on $q$
vertices satisfying \eqref{eq:local-sparsity}.
\end{lemma}

\begin{proof}
Use the configuration model for 3-regular graphs on $q$ labeled vertices.  Fix
$A\subseteq U$ of size $a$, put $\alpha=17/16$, and let
$k=\lfloor\alpha a\rfloor+1$.  If $e_G(A)\ge k$, then some $2k$ of the $3a$
half-edges incident to $A$ are paired among themselves.  A union bound over
the choice and pairing of these half-edges gives
\begin{equation}
  \Prob(e_G(A)\ge k)
  \le
  \binom{3a}{2k}(2k-1)!!
  \prod_{i=0}^{k-1}\frac{1}{3q-(2i+1)}.
  \label{eq:fixed-set-pairing}
\end{equation}
Choose $\beta<1/(2\alpha)$.  If $a\le\beta q$ and $q$ is sufficiently large,
every denominator in \eqref{eq:fixed-set-pairing} is at least $q$.  Using
$k!\ge(k/\eexp)^k$ and $k\ge\alpha a$, there is an absolute constant $C>0$
such that
\begin{equation}
  \Prob(e_G(A)>\alpha a)
  \le\left(\frac{Ca}{q}\right)^k.
  \label{eq:fixed-set-local-sparsity}
\end{equation}
After decreasing $\beta$ so that $C\beta\le1$, a union bound over all sets of
size $a$ yields
\begin{align*}
 &\Prob\bigl(\exists A\subseteq U:|A|=a,
                 \ e_G(A)>\alpha a\bigr)\\
 &\qquad\le
 \binom{q}{a}\left(\frac{Ca}{q}\right)^{\alpha a}
 \le
 \left[\eexp C^\alpha
       \left(\frac{a}{q}\right)^{\alpha-1}\right]^a.
\end{align*}
Taking $\beta$ sufficiently small and summing over
$a\le\sqrt q$ and $\sqrt q<a\le\beta q$ shows that
\eqref{eq:local-sparsity} fails with probability $o(1)$.

For 3-regular configurations, the probability of simplicity is bounded below by
a positive constant independent of $q$~\cite{Janson2009}.  Conditioned on
simplicity, a uniformly random simple 3-regular graph is connected with
probability $1-o(1)$~\cite{Wormald1981}.  Thus simplicity, connectivity, and
\eqref{eq:local-sparsity} occur simultaneously with positive probability for
all sufficiently large even $q$.
\end{proof}

On such a graph, let $\xi_t\subseteq U$ be the threshold contact process with
threshold~$1$ and infection rate $\lambda=1$: every infected vertex recovers
at rate one, and every healthy vertex with at least one infected neighbor
becomes infected at rate one.  Write
\[
  \tau_{\mathrm{cp}}=\inf\{t\ge0:\xi_t=\varnothing\}.
\]

\begin{lemma}[Positive drift of the potential]
\label{lem:contact-drift}
Assume that $G$ satisfies \eqref{eq:local-sparsity}.  For a nonempty infected
set $A$, write
\[
  a=|A|,
  \qquad
  e=e_G(A),
  \qquad
  B=N_G(A)\setminus A,
  \qquad
  b=|B|,
\]
and let $r=e_G(A,U\setminus A)=3a-2e$.  If $a\le\beta q/4$, then
\begin{equation}
  b\ge\frac{31a-16e}{17}.
  \label{eq:healthy-boundary}
\end{equation}
For
\[
  \Psi(A)=|A|+\frac{2}{11}e_G(A,U\setminus A),
\]
the generator $L$ of the contact process satisfies
\begin{equation}
  L\Psi(A)\ge\frac{2}{11}|A|.
  \label{eq:contact-positive-drift}
\end{equation}
\end{lemma}

\begin{proof}
Every vertex of $B$ is incident to an edge from $A$, so $b\le r\le3a$.
Hence $|A\cup B|\le4a\le\beta q$.  The induced graph on $A\cup B$ contains
all $e$ edges inside $A$ and all $r$ edges from $A$ to $B$.  By
\eqref{eq:local-sparsity},
\[
  3a-e=e+r\le e_G(A\cup B)
  \le\frac{17}{16}(a+b),
\]
which is equivalent to \eqref{eq:healthy-boundary}.

The total birth and death rates are $b$ and $a$, so the drift of $|A|$ is
$b-a$.  If a healthy vertex with $s\in\{1,2,3\}$ infected neighbors becomes
infected, the edge boundary changes by $3-2s$; summing over birth vertices
gives $3b-2r$.  If an infected vertex with $s\in\{0,1,2,3\}$ infected
neighbors recovers, the boundary changes by $2s-3$; summing over infected
vertices gives $4e-3a$.  Therefore
\[
  Lr=3b-2r+4e-3a=3b-9a+8e.
\]
It follows that
\begin{align*}
  L\Psi(A)
  &=\frac{17}{11}b+\frac{16}{11}e-\frac{29}{11}a\\
  &\ge
    \frac{31a-16e}{11}
    +\frac{16}{11}e-\frac{29}{11}a
   =\frac{2}{11}a,
\end{align*}
where the inequality uses \eqref{eq:healthy-boundary}.
\end{proof}

\begin{lemma}[Exponential survival of the threshold contact process]
\label{lem:contact-survival}
There are absolute constants $c_0,c_1>0$ such that, for every sufficiently
large even integer $q$, there exists a connected simple 3-regular graph $G_q$ for
which the threshold contact process with threshold~$1$ and $\lambda=1$,
started from the product $\operatorname{Bernoulli}(1/2)$ distribution,
satisfies
\begin{equation}
  \Prob\bigl(\tau_{\mathrm{cp}}\le\eexp^{c_0q}\bigr)
  \le\eexp^{-c_1q}.
  \label{eq:contact-survival-tail}
\end{equation}
\end{lemma}

\begin{proof}
Choose $G_q$ from \Cref{lem:locally-sparse-cubic}, decrease $\beta$ if
necessary so that $\beta/4<1/2$, and put
\[
  h=\left\lfloor\frac{\beta q}{4}\right\rfloor.
\]
Let $A_0,A_1,\ldots$ be the embedded jump chain of the contact process.  At a
state $A$ with $a=|A|$, the total transition rate is
\[
  R(A)=a+|N_G(A)\setminus A|\le4a.
\]
Thus \eqref{eq:contact-positive-drift} implies
\begin{equation}
  \E{\Psi(A_{j+1})-\Psi(A_j)\mid A_j=A}
  \ge\frac{1}{22}
  \label{eq:embedded-positive-drift}
\end{equation}
whenever $1\le a\le h$.  A single birth or death changes $|A|$ by one and the
edge boundary by at most three, so
\begin{equation}
  |\Psi(A_{j+1})-\Psi(A_j)|\le\frac{17}{11}.
  \label{eq:contact-bounded-jump}
\end{equation}

Set $\delta=1/22$, $K=17/11$, and $s=\delta/K^2$.  Conditional Hoeffding's
lemma gives, whenever $1\le|A_j|\le h$,
\[
  \E{\exp\bigl(-s(\Psi(A_{j+1})-\Psi(A_j))\bigr)\mid A_j}
  \le \exp\left(-\frac{s\delta}{2}\right)<1.
\]
Start the jump chain from a state with $|A_0|=h$ and stop it on first hitting
size $0$ or $h+1$.  The stopped process $\exp(-s\Psi(A_j))$ is a nonnegative
supermartingale.  Applying optional stopping to deterministic truncations and
then Fatou's lemma gives
\begin{equation}
  \Prob_{A_0}\bigl(\text{hit size $0$ before size $h+1$}\bigr)
  \le\eexp^{-sh}.
  \label{eq:one-low-density-excursion}
\end{equation}

Under the product $\operatorname{Bernoulli}(1/2)$ initial distribution, a
Chernoff bound gives a constant $c_I>0$ such that
\begin{equation}
  \Prob(|A_0|\le h)\le\eexp^{-c_Iq}.
  \label{eq:contact-initial-density}
\end{equation}
Let $J=\lfloor\eexp^{sh/2}\rfloor$.  On $|A_0|>h$, every path to the empty set
must first enter level $h$.  At each entrance to level $h$, the strong Markov
property and \eqref{eq:one-low-density-excursion} bound the conditional
probability of reaching zero before returning to level $h+1$ by
$\eexp^{-sh}$.  A union bound over at most $J$ such entrances during the first
$J$ jumps yields
\begin{equation}
  \Prob(\text{extinction during the first $J$ jumps})
  \le\eexp^{-c_Iq}+J\eexp^{-sh}
  \le\eexp^{-c_Iq}+\eexp^{-sh/2}.
  \label{eq:contact-first-J-jumps}
\end{equation}

At every state the total transition rate is at most $q$.  By uniformization,
the number of jumps by time $t$ is stochastically dominated by a Poisson
random variable with mean $qt$.  Take $t=\eexp^{sh/4}$.  For all sufficiently
large $q$,
\[
  \Prob(\operatorname{Pois}(qt)\ge J)
  \le\left(\frac{\eexp qt}{J}\right)^J
  \le2^{-J}.
\]
Combining this with \eqref{eq:contact-first-J-jumps} proves
\eqref{eq:contact-survival-tail}, after adjusting the absolute constants.
\end{proof}

\subsection{Completing the MT lower bound}

We couple the threshold contact process below the set of vertices in either
voter state and use the survival bound to control the consensus time.

\begin{lemma}[Exponential voter consensus time]
\label{lem:voter-consensus-time}
For the graph family in \Cref{lem:contact-survival}, there are absolute
constants $c,c'>0$ such that the threshold voter model with threshold~$1$,
started from the product $\operatorname{Bernoulli}(1/2)$ distribution,
satisfies
\begin{equation}
  \Prob(\tau_{\mathrm{cons}}>\eexp^{cq})
  \ge1-\eexp^{-c'q}
  \label{eq:voter-consensus-tail}
\end{equation}
for all sufficiently large $q$.
\end{lemma}

\begin{proof}
At each vertex, take independent rate-one up-clocks and down-clocks.  For the
voter model, an up-clock changes state $0$ to state $1$ if the vertex has a
neighbor in state $1$, and a down-clock changes state $1$ to state $0$ if the
vertex has a neighbor in state $0$.  This is the threshold voter model with
threshold~$1$.

Using the same clocks, construct a threshold contact process $\xi_t^1$ from
the initial set of vertices in voter state $1$: an up-clock infects a healthy
vertex if it has an infected neighbor, and a down-clock recovers the vertex if
it is infected.  Induction over clock events gives
\[
  \xi_t^1\subseteq\{v:\eta_t(v)=1\}
  \qquad\text{for all }t\ge0.
\]
Hence consensus at all zero by time $t$ implies extinction of $\xi^1$ by time
$t$.  Applying the same construction to the complement of the voter
configuration gives a contact process $\xi_t^0$ such that consensus at all one
implies extinction of $\xi^0$.  Both contact processes have the product
$\operatorname{Bernoulli}(1/2)$ initial distribution.  Therefore
\Cref{lem:contact-survival} and a union bound give
\[
  \Prob(\tau_{\mathrm{cons}}\le\eexp^{c_0q})
  \le2\eexp^{-c_1q}.
\]
After adjusting constants, this is \eqref{eq:voter-consensus-tail}.
\end{proof}

Now we can prove Theorem~\ref{thm-mt-exponential}.

\begin{proof}[Proof of Theorem~\ref{thm-mt-exponential}]
Fix a sufficiently large integer $m$.  Let
\[
  q=2\left\lfloor\frac{m}{8}\right\rfloor,
  \qquad
  r=m-4q.
\]
Then $q$ is even, $0\le r\le7$, and $q=\Theta(m)$.  Let $G_q$ be the graph
from \Cref{lem:contact-survival}, and let $F^\circ=F(G_q)$.  The formula
$F^\circ$ has $4q$ clauses.  Introduce $3r$ fresh variables
$a_i,b_i,c_i$ and set
\[
  P_r=\bigwedge_{i=1}^r(a_i\vee b_i\vee c_i),
  \qquad
  F_m=F^\circ\wedge P_r.
\]
Thus $F_m$ has exactly $m$ clauses, every clause contains three distinct
variables, $F_m$ is satisfiable, and $\Delta(F_m)=8$.

First consider WalkSAT.  The variables of $P_r$ do not occur in $F^\circ$.
Conditional on a step selecting a clause of $F^\circ$, that clause is uniform
among the currently unsatisfied clauses of $F^\circ$; hence the subsequence of
updates to $F^\circ$ has exactly the law of WalkSAT on $F^\circ$.  Moreover, a
padding clause can be selected at most once: if it is unsatisfied, all three of
its variables are zero, and the selected flip makes that clause satisfied
permanently.  Therefore
\[
  \E{T_{\WS}(F_m)}
  \le \E{T_{\WS}(F^\circ)}+r
  =O(q^4)
  =O(m^4)
\]
by \Cref{cor:tseitin-ws-bound}.

For MT, the subsequence of resamplings that select a clause of $F^\circ$ has
exactly the law of MT on $F^\circ$.  The total number of MT steps on $F_m$ is
therefore at least the number of such base-formula resamplings required to
satisfy $F^\circ$.  By \Cref{lem:mt-voter-times,lem:voter-consensus-time},
under the uniform initialization,
\[
  \Prob\bigl(\tau_\varnothing>2\eexp^{cq}\bigr)
  \ge1-\eexp^{-c'q}.
\]
Hence
\[
  \E{\tau_\varnothing}
  \ge2\eexp^{cq}\bigl(1-\eexp^{-c'q}\bigr)
  =\eexp^{\Omega(q)}.
\]
Using \eqref{eq:mt-time-comparison},
\[
  \E{T_{\MT}(F^\circ)}
  \ge2\E{\tau_\varnothing}
  =\eexp^{\Omega(q)}.
\]
Consequently,
\[
  \E{T_{\MT}(F_m)}
  \ge \E{T_{\MT}(F^\circ)}
  =\eexp^{\Omega(q)}
  =2^{\Omega(m)}.
\]
This proves the theorem for every sufficiently large $m$.
\end{proof}

\begin{remark}
The construction in this section has maximum variable occurrence eight.  It is
used for the WalkSAT--MT separation; the occurrence-three threshold is
established separately in \Cref{sec:joint-hard-family}.
\end{remark}

\section{A 3-CNF Family with Exponential WalkSAT and Polynomial MT Time}
\label{sec-ws-slow}

In this section, we prove \Cref{thm-ws-exponential} by constructing a family of 3-CNF formulas based on expander graphs.
For this family, \Cref{prop:direct-mt-time} establishes a linear expected-time bound for the MT algorithm, while \Cref{prop:direct-ws-lower} establishes an exponential lower bound on the expected running time of WalkSAT.
The theorem follows immediately from these two results.

\subsection{Constructing Formulas from Expander Graphs}
\label{subsec:direct-construction}

For a graph $G=(V,E)$, let
\[
  \lambda_1(G)\ge\lambda_2(G)\ge\cdots\ge\lambda_{|V|}(G)
\]
be the eigenvalues of the adjacency matrix of $G$. 
For distinct odd primes $p$ and $q$, we write
$\left(\frac{p}{q}\right)$ for the Legendre symbol.
We recall the LPS
construction~\cite{LPS1988} of expander graphs as follows.

\begin{theorem}[{\cite[Theorems~3.4 and~4.1]{LPS1988}}]
\label{thm:lps-graphs}
Let $p$ and $q$ be distinct primes satisfying
$q>p$, $p\equiv q\equiv1\pmod4$, and
$\left(\frac{p}{q}\right)=1$.  The LPS construction gives an explicit
simple connected $(p+1)$-regular graph $X^{p,q}$ satisfying
\[
  |V(X^{p,q})|=\frac{q(q^2-1)}2,
  \qquad
  \lambda_1(X^{p,q})=p+1,
\]
and
\[
  \max_{2\le i\le |V(X^{p,q})|}|\lambda_i(X^{p,q})|
  \le2\sqrt p.
\]
\end{theorem}

\begin{lemma}
\label{lem:explicit-expander-input}
There exists an explicitly constructible infinite family of simple connected
$102$-regular graphs $\{G_j=(V_j,E_j)\}_{j\ge1}$
such that $|V_j|\to\infty$ as $j\to\infty$ and, for every $j\ge1$,
\[
  \lambda_1(G_j)=102,
  \qquad
  \max_{2\le i\le |V_j|}|\lambda_i(G_j)|<21.
\]
\end{lemma}

\begin{proof}
By Dirichlet's theorem, there are infinitely many primes
\[
  q_1<q_2<\cdots,
  \qquad
  q_j\equiv1\pmod{404}.
\]
In particular, $q_j>101$, $q_j\equiv1\pmod4$, and
$q_j\equiv1\pmod{101}$.  Since also $101\equiv1\pmod4$, quadratic
reciprocity gives
\[
  \left(\frac{101}{q_j}\right)
  =\left(\frac{q_j}{101}\right)
  =\left(\frac{1}{101}\right)
  =1.
\]
Apply Theorem~\ref{thm:lps-graphs} with $p=101$ and $q=q_j$, and set
$G_j=X^{101,q_j}$.  The theorem shows that $G_j$ is simple, connected, and
$102$-regular, with
\[
  |V_j|=\frac{q_j(q_j^2-1)}2.
\]
Since $q_j$ tends to infinity, so does $|V_j|$.  The same theorem gives
\[
  \lambda_1(G_j)=102,
  \qquad
  \max_{2\le i\le |V_j|}|\lambda_i(G_j)|
  \le2\sqrt{101}<21.
\]
The LPS construction and the increasing enumeration of the primes $q_j$ are
explicit, which completes the proof.
\end{proof}

\begin{lemma}
\label{lem:spectral-cut}
Let $G=(V,E)$ be a graph from Lemma~\ref{lem:explicit-expander-input}, and let
$Z\subseteq V$ satisfy $|Z|\le |V|/4$.  Then, for $S=V\setminus Z$,
\[
  e_G(S,Z)\ge\frac{243}{4}|Z|.
\]
\end{lemma}

\begin{proof}
Let $A$ be the adjacency matrix of $G$ and write
$\one_Z=(|Z|/|V|)\one+f$, where $f\perp\one$.  Then
$\|f\|_2^2=|Z|(1-|Z|/|V|)$, and regularity gives
\[
  e_G(S,Z)
  =102|Z|-\one_Z^{\top}A\one_Z
  =102|Z|-\frac{102|Z|^2}{|V|}-f^{\top}Af.
\]
Since $A$ is symmetric, the spectral theorem and
Lemma~\ref{lem:explicit-expander-input} give
$f^{\top}Af\le21\|f\|_2^2$.  Hence
\[
  e_G(S,Z)
  \ge81|Z|\left(1-\frac{|Z|}{|V|}\right)
  \ge\frac{243}{4}|Z|,
\]
where the last inequality uses $|Z|\le |V|/4$.
\end{proof}

\paragraph{Direct clause construction.}
Fix a graph $G=(V,E)$ from Lemma~\ref{lem:explicit-expander-input}.  We define
$F(G)$ as follows, regarding each undirected edge as two directed edges.  Each
vertex $v$ carries two progress variables $p_v,q_v$, and its block is active
when $(p_v,q_v)=(1,1)$.

For every $v\in V$, $r\in\{p,q\}$, and $c\in\{1,\ldots,52\}$, introduce
fresh variables $a_{v,r,c},b_{v,r,c}$ and add the four-clause gadget
\[
\begin{aligned}
 &(r_v\lor a_{v,r,c}\lor b_{v,r,c}),
 &&(r_v\lor a_{v,r,c}\lor\neg b_{v,r,c}),\\
 &(r_v\lor\neg a_{v,r,c}\lor b_{v,r,c}),
 &&(r_v\lor\neg a_{v,r,c}\lor\neg b_{v,r,c}).
\end{aligned}
\]
Exactly one of these four clauses is unsatisfied when $r_v=0$, and none is
unsatisfied when $r_v=1$.

For every $v\in V$ and $c\in\{1,\ldots,613\}$, introduce a fresh variable
$h_{v,c}$ and add the two-clause gadget
\[
  (p_v\lor q_v\lor h_{v,c}),
  \qquad
  (p_v\lor q_v\lor\neg h_{v,c}).
\]
Exactly one clause in this pair is unsatisfied precisely when
$(p_v,q_v)=(0,0)$.

Finally, for every directed edge $u\to v$, add
\[
  (p_v\lor\neg p_u\lor\neg q_u),
  \qquad
  (q_v\lor\neg p_u\lor\neg q_u).
\]
These clauses are unsatisfied precisely when $u$ is active and the
corresponding target bit at $v$ is zero.  The conjunction of all clauses above
is $F(G)$.

Conceptually, the four-clause gadget pads the unit constraint $r_v$, the
two-clause gadget pads $p_v\lor q_v$, and the edge clauses express
$(p_u\land q_u)\Rightarrow p_v,q_v$.  The fresh variables make every clause a
proper 3-clause without changing these logical constraints.

For an assignment $x$, partition the vertices into
\[
\begin{aligned}
  S(x)&=\{v:(p_v,q_v)=(1,1)\},\\
  Z_1(x)&=\{v:(p_v,q_v)\in\{(1,0),(0,1)\}\},\\
  Z_0(x)&=\{v:(p_v,q_v)=(0,0)\}.
\end{aligned}
\]

We summarize the properties of the formula constructed above as follows.
\begin{lemma}
\label{lem:direct-structural-audit}
The formula $F(G)$ constructed above has following properties:
\begin{itemize}
  \item Every clause of $F(G)$ contains three distinct variables and is
non-tautological. 
  \item  The formula has $823|V|$ variables and $1846|V|$ clauses.
Every variable appears in at most $1740$ clauses.
  \item 
For every assignment $x$, the number $B(x)$ of unsatisfied clauses is
\[
  B(x)=717|Z_0(x)|+52|Z_1(x)|
       +2e_G(S(x),Z_0(x))+e_G(S(x),Z_1(x)).
\]
In particular, $x$ satisfies $F(G)$ if and only if $S(x)=V$.
\end{itemize}
\end{lemma}

\begin{proof}
Fresh auxiliary variables ensure that every clause in a four- or two-clause
gadget contains three distinct variables.  An edge clause uses one progress
variable at its target $v$ and both progress variables at its source $u$.
These are distinct because $G$ is simple.  No displayed clause is tautological.

At each vertex, the four-clause gadgets introduce $2\cdot52\cdot2=208$
auxiliary variables and $2\cdot52\cdot4=416$ clauses.  The two-clause gadgets
introduce $613$ variables and $1226$ clauses.  The $102|V|$ directed edges add
$204|V|$ clauses.  Hence
\[
  (2+208+613)|V|=823|V|,
  \qquad
  (416+1226+204)|V|=1846|V|.
\]

A progress variable occurs $208$ times in the four-clause gadgets and $1226$
times in the two-clause gadgets.  It also occurs in $102$ edge clauses as a
target variable and $204$ as a source variable, for a total of $1740$.  Each
$a$- or $b$-variable occurs four times, and each $h$-variable twice.

Fix an assignment $x$.  A zero progress bit makes exactly one clause in each
of its $52$ four-clause gadgets unsatisfied.  Thus each vertex in $Z_1(x)$
accounts for $52$ unsatisfied clauses.  A vertex in $Z_0(x)$ has two zero bits
and also makes one clause in each of its $613$ two-clause gadgets unsatisfied,
for a total of $2\cdot52+613=717$.

An edge clause for $u\to v$ is unsatisfied exactly when $u\in S(x)$ and its
target bit at $v$ is zero.  Consequently, an edge between $S(x)$ and $Z_1(x)$
accounts for one unsatisfied clause, while an edge between $S(x)$ and $Z_0(x)$
accounts for two.  Adding these counts gives the formula for $B(x)$.  Since
$Z_0(x)\sqcup Z_1(x)=V\setminus S(x)$, we have $B(x)=0$ exactly when $S(x)=V$.
\end{proof}

\subsection{Linear time for uniform-clause Moser--Tardos}

Run uniform-clause Moser--Tardos on $F(G)$.  For each $t$, write
\[
  S_t=S(X_t),
  \qquad Z_{0,t}=Z_0(X_t),
  \qquad Z_{1,t}=Z_1(X_t).
\]
The process terminates exactly when $S_t=V$.  We first show a uniform positive
drift before termination.

\begin{lemma}
\label{lem:direct-mt-drift}
Suppose that the current assignment $X_t$ does not satisfy $F(G)$.  Recall
from Lemma~\ref{lem:direct-structural-audit} that $B(X_t)$ is the number of
clauses unsatisfied by $X_t$.  Then
\begin{equation}
  \E{|S_{t+1}|-|S_t|\mid\mathcal F_t}
  =\frac{104|Z_{1,t}|+613|Z_{0,t}|-e_G(S_t,Z_{1,t})-6e_G(S_t,Z_{0,t})}
  {4B(X_t)}.
\label{eq:direct-mt-exact-drift}
\end{equation}
In particular,
\begin{equation}
  \E{|S_{t+1}|-|S_t|\mid\mathcal F_t}\ge\frac1{3684}.
  \label{eq:direct-mt-uniform-drift}
\end{equation}
\end{lemma}

\begin{proof}
Let $C_t$ denote the unsatisfied clause selected at time $t$.  Condition on
$\mathcal F_t$, so the current assignment $X_t$ is fixed.

First take $v\in Z_{1,t}$ with
$(p_v,q_v)=(0,1)$.  Fix one of the $52$ four-clause gadgets associated with
$p_v$, let $C$ be its unique unsatisfied clause, and suppose $C_t=C$.

Resampling $C$ redraws $p_v$ and the two auxiliary variables in $C$
independently and uniformly.  The value of $q_v$ remains $1$, and all other
progress variables remain unchanged.  Hence $|S_{t+1}|-|S_t|$ equals $1$ if
$p_v$ is redrawn to $1$, and equals $0$ otherwise.  Therefore
\[
  \E{|S_{t+1}|-|S_t|\mid\mathcal F_t,C_t=C}
  =1\cdot\frac12+0\cdot\frac12=\frac12.
\]

The case $(p_v,q_v)=(1,0)$ is identical after interchanging $p_v$ and $q_v$.

Next take $v\in Z_{0,t}$, so $(p_v,q_v)=(0,0)$, and let $C$ be the unique
unsatisfied clause in one of the four-clause gadgets associated with $p_v$.
Suppose $C_t=C$.  Resampling $C$ redraws $p_v$, but $q_v$ remains $0$.
Thus $v$ is inactive both before and after the step, and
\[
  \E{|S_{t+1}|-|S_t|\mid\mathcal F_t,C_t=C}=0.
\]
The same argument applies to a four-clause gadget associated with $q_v$.

Still take $(p_v,q_v)=(0,0)$, but now let $C$ be the unique unsatisfied clause
in one of the two-clause gadgets at $v$, and suppose $C_t=C$.  Resampling $C$
redraws $p_v$ and $q_v$ independently.  The vertex $v$ becomes active exactly
when both are redrawn to $1$, which has probability $1/4$.  Hence
\[
  \E{|S_{t+1}|-|S_t|\mid\mathcal F_t,C_t=C}
  =1\cdot\frac14+0\cdot\frac34=\frac14.
\]

Now take a directed edge $u\to v$ with $u\in S_t$ and $v\in Z_{1,t}$.
Assume $(p_v,q_v)=(0,1)$ and let
$C=(p_v\lor\neg p_u\lor\neg q_u)$, which is unsatisfied, and suppose
$C_t=C$.  Resampling $C$ redraws $p_v,p_u,q_u$ independently, while $q_v$
remains $1$.  Before the step, $u$ is active and $v$ is inactive.  Afterward,
$u$ is active with probability $1/4$ and $v$ with probability $1/2$.
Therefore
\[
  \E{|S_{t+1}|-|S_t|\mid\mathcal F_t,C_t=C}
  =\frac14+\frac12-1=-\frac14.
\]
The case $(p_v,q_v)=(1,0)$ is symmetric.

Finally, take $u\in S_t$ and $v\in Z_{0,t}$, and again let
$C=(p_v\lor\neg p_u\lor\neg q_u)$.  This clause is unsatisfied; suppose
$C_t=C$.  Before the step, only $u$ is active.  Resampling $C$ leaves
$q_v=0$, so $v$ remains inactive, while $u$ remains active with probability
$1/4$.  Thus
\[
  \E{|S_{t+1}|-|S_t|\mid\mathcal F_t,C_t=C}
  =\frac14-1=-\frac34.
\]
The edge clause containing $q_v$ gives the same value.

These five calculations are summarized below.  For each row, the last column
gives the conditional expectation for any clause $C$ of the indicated family
and location.
\begin{center}
\begin{tabular}{lll}
  \toprule
  Clause family & Location & Value \\
  \midrule
  four-clause gadget & $Z_{1,t}$ & $+1/2$ \\
  four-clause gadget & $Z_{0,t}$ & $0$ \\
  two-clause gadget  & $Z_{0,t}$ & $+1/4$ \\
  edge clause        & target in $Z_{1,t}$ & $-1/4$ \\
  edge clause        & target in $Z_{0,t}$ & $-3/4$ \\
  \bottomrule
\end{tabular}
\end{center}

There are $52|Z_{1,t}|$ unsatisfied clauses from four-clause gadgets at
$Z_{1,t}$ blocks, $104|Z_{0,t}|$ at $Z_{0,t}$ blocks, and $613|Z_{0,t}|$
unsatisfied clauses from two-clause gadgets.  There are also
$e_G(S_t,Z_{1,t})$ unsatisfied edge clauses with target in $Z_{1,t}$ and
$2e_G(S_t,Z_{0,t})$ with target in $Z_{0,t}$.  Since $C_t$ is uniform among
these $B(X_t)$ clauses, averaging the table values over $C_t$ proves
\eqref{eq:direct-mt-exact-drift}.

Since $G$ is $102$-regular,
$e_G(S_t,Z_{i,t})\le102|Z_{i,t}|$ for $i\in\{0,1\}$.  The numerator in
\eqref{eq:direct-mt-exact-drift} is therefore at least
\[
  \left(26-\frac{102}{4}\right)|Z_{1,t}|
  +\left(\frac{613}{4}-\frac{3\cdot102}{2}\right)|Z_{0,t}|
  =\frac{2|Z_{1,t}|+|Z_{0,t}|}{4}
  \ge\frac{|Z_{0,t}|+|Z_{1,t}|}{4}.
\]
Lemma~\ref{lem:direct-structural-audit} also gives
\[
  B(X_t)\le(717+2\cdot102)|Z_{0,t}|+(52+102)|Z_{1,t}|
  =921|Z_{0,t}|+154|Z_{1,t}|
  \le921(|Z_{0,t}|+|Z_{1,t}|).
\]
Because $X_t$ does not satisfy $F(G)$,
$|Z_{0,t}|+|Z_{1,t}|>0$.  Dividing these two bounds proves
\eqref{eq:direct-mt-uniform-drift}.
\end{proof}

\begin{proposition}[Moser--Tardos hitting-time bounds]
\label{prop:direct-mt-time}
For every initial assignment $X_0$,
\begin{equation}
  \E{T_{\MT}\mid X_0}\le3684\bigl(|V|-|S_0|\bigr).
  \label{eq:direct-mt-initial-bound}
\end{equation}
Under independent uniform initialization,
\begin{equation}
  \frac{3|V|}{4}\le\E{T_{\MT}}\le2763|V|,
  \qquad
  \Prob(T_{\MT}>7368|V|)
  \le\exp\left(-\frac{|V|}{58944}\right).
  \label{eq:direct-mt-time-bounds}
\end{equation}
\end{proposition}

\begin{proof}
Freeze the process after termination and set $T_n=T_{\MT}\wedge n$.  By
Lemma~\ref{lem:direct-mt-drift}, summing the conditional drift before $T_n$
gives
\[
  \E{|S_{T_n}|-|S_0|}\ge\frac1{3684}\E{T_n}.
\]
Since $|S_{T_n}|\le|V|$, we obtain
$\E{T_n}\le3684(|V|-|S_0|)$.  Monotone convergence proves
\eqref{eq:direct-mt-initial-bound} and the integrability of $T_{\MT}$.

Under uniform initialization, the blocks are independently active with
probability $1/4$.  Hence
\[
  |S_0|\sim\operatorname{Bin}(|V|,1/4),
  \qquad \E{|S_0|}=|V|/4.
\]
Averaging \eqref{eq:direct-mt-initial-bound} gives the upper bound
$2763|V|$.  Resampling a clause from a four-clause or two-clause gadget
affects only one block.  For an edge clause, the source can deactivate and the
target can activate, but these changes have opposite signs.  Hence
$\bigl||S_{t+1}|-|S_t|\bigr|\le1$, so reaching $S_t=V$ requires at least
$|V|-|S_0|$ steps.  Taking expectations gives the lower bound $3|V|/4$.

For the tail bound, define
\[
  \mu_t=\E{|S_{t+1}|-|S_t|\mid\mathcal F_t},
  \qquad
  M_t=\sum_{s=0}^{t-1}
  \bigl(|S_{s+1}|-|S_s|-\mu_s\bigr)
\]
for the frozen process.  Since
$\bigl||S_{t+1}|-|S_t|\bigr|\le1$ and
$|\mu_t|\le1$, every martingale increment of $M_t$ has absolute value at most
two.

Set $t_*=7368|V|$.  On $\{T_{\MT}>t_*\}$, the algorithm has not terminated at
any time $s<t_*$.  Lemma~\ref{lem:direct-mt-drift} therefore gives
\[
  \sum_{s<t_*}\mu_s\ge\frac{t_*}{3684}=2|V|.
\]
Because $|S_{t_*}|-|S_0|\le|V|$, this event implies
$M_{t_*}\le-|V|$.
Azuma--Hoeffding therefore yields
\[
  \Prob(T_{\MT}>t_*)
  \le\exp\left(-\frac{|V|^2}{8t_*}\right)
  =\exp\left(-\frac{|V|}{58944}\right),
\]
which proves \eqref{eq:direct-mt-time-bounds}.
\end{proof}

\subsection{WalkSAT: an exponential barrier}

Run uniform-clause WalkSAT on $F(G)$.  In this subsection, $X_t$ denotes
its assignment at time $t$, and we again write
\[
  S_t=S(X_t),
  \qquad Z_{0,t}=Z_0(X_t),
  \qquad Z_{1,t}=Z_1(X_t).
\]
Set
\[
  H=\left\lceil\frac{3|V|}{4}\right\rceil,
  \qquad \rho=\frac{451}{486}.
\]
We first show that, once $|S_t|\ge H$, a downward move is uniformly more
likely than an upward move.

\begin{lemma}[High-progress backward bias]
\label{lem:direct-ws-bias}
For $H\le |S_t|<|V|$, let
\[
  p_t^\pm
  =\Prob\bigl(|S_{t+1}|-|S_t|=\pm1\mid\mathcal F_t\bigr).
\]
Recall from Lemma~\ref{lem:direct-structural-audit} that $B(X_t)$ is the number
of clauses unsatisfied by $X_t$.  Then
\begin{equation}
  p_t^+=\frac{52|Z_{1,t}|+e_G(S_t,Z_{1,t})}{3B(X_t)},
  \qquad
  p_t^-=\frac{2e_G(S_t,Z_{1,t})+4e_G(S_t,Z_{0,t})}{3B(X_t)},
  \label{eq:direct-ws-transitions}
\end{equation}
and
\begin{equation}
  p_t^+\le\rho p_t^-.
  \label{eq:direct-ws-bias}
\end{equation}
\end{lemma}

\begin{proof}
Let $C_t$ denote the unsatisfied clause selected at time $t$, and write
$\Delta_t=|S_{t+1}|-|S_t|$.  Condition on $\mathcal F_t$, so $X_t$ is fixed.
Given $C_t$, WalkSAT chooses one of its three variables uniformly and
flips that variable.

First take $v\in Z_{1,t}$ with $(p_v,q_v)=(0,1)$.  Let $C$ be the unique
unsatisfied clause in one of the four-clause gadgets associated with $p_v$,
and suppose $C_t=C$.  Flipping $p_v$ activates $v$, whereas flipping either
auxiliary variable leaves every block unchanged.  Hence
\[
  \Prob(\Delta_t=1\mid\mathcal F_t,C_t=C)=\frac13,
  \qquad
  \Prob(\Delta_t=-1\mid\mathcal F_t,C_t=C)=0.
\]
The case $(p_v,q_v)=(1,0)$ is symmetric.

Next take $v\in Z_{0,t}$, so $(p_v,q_v)=(0,0)$, and let $C$ be the unique
unsatisfied clause in a four-clause gadget associated with $p_v$.  Suppose
$C_t=C$.  If $p_v$ is flipped, then $(p_v,q_v)$ becomes $(1,0)$; either
auxiliary flip leaves it at $(0,0)$.  In both cases $v$ remains inactive, so
\[
  \Prob(\Delta_t=1\mid\mathcal F_t,C_t=C)
  =\Prob(\Delta_t=-1\mid\mathcal F_t,C_t=C)=0.
\]
The same argument applies to a gadget associated with $q_v$.

Still take $(p_v,q_v)=(0,0)$, but now let $C$ be the unique unsatisfied clause
in a two-clause gadget at $v$, and suppose $C_t=C$.  Flipping $p_v$ or $q_v$
changes the pair to $(1,0)$ or $(0,1)$, and flipping the auxiliary variable
leaves it at $(0,0)$.  Thus $v$ remains inactive in all three cases, and
\[
  \Prob(\Delta_t=1\mid\mathcal F_t,C_t=C)
  =\Prob(\Delta_t=-1\mid\mathcal F_t,C_t=C)=0.
\]

Now take a directed edge $u\to v$ with $u\in S_t$ and $v\in Z_{1,t}$.
Assume $(p_v,q_v)=(0,1)$ and let
$C=(p_v\lor\neg p_u\lor\neg q_u)$, which is unsatisfied; suppose $C_t=C$.
Flipping $p_v$ activates $v$ and leaves $u$ active, so $\Delta_t=1$.
Flipping $p_u$ or $q_u$ deactivates $u$ and leaves $v$ inactive, so
$\Delta_t=-1$.  Therefore
\[
  \Prob(\Delta_t=1\mid\mathcal F_t,C_t=C)=\frac13,
  \qquad
  \Prob(\Delta_t=-1\mid\mathcal F_t,C_t=C)=\frac23.
\]
The case $(p_v,q_v)=(1,0)$ is symmetric.

Finally, take $u\in S_t$ and $v\in Z_{0,t}$, and again let
$C=(p_v\lor\neg p_u\lor\neg q_u)$, which is unsatisfied, and suppose
$C_t=C$.  Flipping $p_v$ changes the target pair to $(1,0)$, so $u$ stays
active and $v$ stays inactive; hence $\Delta_t=0$.  Flipping either source
variable deactivates $u$, so $\Delta_t=-1$.  Hence
\[
  \Prob(\Delta_t=1\mid\mathcal F_t,C_t=C)=0,
  \qquad
  \Prob(\Delta_t=-1\mid\mathcal F_t,C_t=C)=\frac23.
\]
The edge clause containing $q_v$ gives the same probabilities.

These five calculations are summarized below.  The last two columns give the
conditional probabilities of $\Delta_t=1$ and $\Delta_t=-1$, respectively.
\begin{center}
\begin{tabular}{llll}
  \toprule
  Clause family & Location & $\Delta_t=1$ & $\Delta_t=-1$ \\
  \midrule
  four-clause gadget & $Z_{1,t}$ & $1/3$ & $0$ \\
  four-clause gadget & $Z_{0,t}$ & $0$ & $0$ \\
  two-clause gadget  & $Z_{0,t}$ & $0$ & $0$ \\
  edge clause        & target in $Z_{1,t}$ & $1/3$ & $2/3$ \\
  edge clause        & target in $Z_{0,t}$ & $0$ & $2/3$ \\
  \bottomrule
\end{tabular}
\end{center}

There are $52|Z_{1,t}|$ unsatisfied four-clause-gadget clauses at $Z_{1,t}$,
$e_G(S_t,Z_{1,t})$ unsatisfied edge clauses with target in $Z_{1,t}$, and
$2e_G(S_t,Z_{0,t})$ with target in $Z_{0,t}$.  Since $C_t$ is uniform among
the $B(X_t)$ unsatisfied clauses, averaging the table entries proves
\eqref{eq:direct-ws-transitions}.

Because $H\le|S_t|<|V|$,
\[
  1\le |Z_{0,t}|+|Z_{1,t}|
  =|V|-|S_t|\le\frac{|V|}{4}.
\]
Lemma~\ref{lem:spectral-cut} therefore gives
\[
  e_G(S_t,Z_{0,t})+e_G(S_t,Z_{1,t})
  \ge\frac{243}{4}\bigl(|Z_{0,t}|+|Z_{1,t}|\bigr).
\]
In particular, $p_t^->0$.  Also,
$|Z_{1,t}|\le|Z_{0,t}|+|Z_{1,t}|$ and
\[
  e_G(S_t,Z_{1,t})+2e_G(S_t,Z_{0,t})
  \ge e_G(S_t,Z_{1,t})+e_G(S_t,Z_{0,t}).
\]
Separating the two terms in the numerator of $p_t^+$ and using
\eqref{eq:direct-ws-transitions}, we obtain
\begin{align*}
  \frac{p_t^+}{p_t^-}
  &=\frac{52|Z_{1,t}|+e_G(S_t,Z_{1,t})}
          {2e_G(S_t,Z_{1,t})+4e_G(S_t,Z_{0,t})}\\
  &=\frac{e_G(S_t,Z_{1,t})}
          {2\bigl(e_G(S_t,Z_{1,t})+2e_G(S_t,Z_{0,t})\bigr)}
    +\frac{26|Z_{1,t}|}
          {e_G(S_t,Z_{1,t})+2e_G(S_t,Z_{0,t})}\\
  &\le\frac12+
  \frac{26(|Z_{0,t}|+|Z_{1,t}|)}
       {e_G(S_t,Z_{0,t})+e_G(S_t,Z_{1,t})}
  \le\frac12+\frac{104}{243}\\
  &=\frac{451}{486}.
\end{align*}
This proves \eqref{eq:direct-ws-bias}.
\end{proof}

The inequality in Lemma~\ref{lem:direct-ws-bias} controls every attempt to
cross from $H$ to $|V|$ before returning to $H-1$.

\begin{proposition}[Uniform one-attempt crossing bound]
\label{prop:direct-ws-one-attempt}
From any initial assignment $X_0$ with $|S_0|=H$,
\begin{equation}
  \Prob(\text{$S_t=V$ before $|S_t|=H-1$})
  \le\rho^{|V|-H}.
  \label{eq:direct-ws-one-attempt}
\end{equation}
\end{proposition}

\begin{proof}
Let $\tau$ be the first time that $|S_t|\in\{H-1,|V|\}$.  Before $\tau$,
Lemma~\ref{lem:direct-ws-bias} gives $p_t^+\le\rho p_t^-$.  Since
$\rho^{-1}-1>0$ and one step changes $|S_t|$ by $-1$, $0$, or $1$,
\begin{align*}
  \E{\rho^{-|S_{t+1}|}\mid\mathcal F_t}
  &=\rho^{-|S_t|}
    \bigl(1+p_t^+(\rho^{-1}-1)+p_t^-(\rho-1)\bigr)\\
  &\le\rho^{-|S_t|}
    \bigl(1+\rho p_t^-(\rho^{-1}-1)+p_t^-(\rho-1)\bigr)\\
  &=\rho^{-|S_t|}.
\end{align*}
The last equality uses
$\rho(\rho^{-1}-1)+(\rho-1)=0$.  Hence
$\rho^{-|S_{t\wedge\tau}|}$ is a nonnegative supermartingale.

The nonnegative-supermartingale maximal inequality, applied up to time $n$,
gives
\[
  \Prob(\text{$S_t=V$ for some $t\le n$ before $|S_t|=H-1$})
  \le\frac{\rho^{-H}}{\rho^{-|V|}}=\rho^{|V|-H}.
\]
Letting $n\to\infty$ proves \eqref{eq:direct-ws-one-attempt}.  The argument
conditions on the complete assignment at every step, so it does not require
the process $(|S_t|)$ to be Markovian.
\end{proof}

Uniform initialization begins below $H$ with exponentially high probability.
Combining this fact with the one-attempt bound gives the desired lower bound.

\begin{proposition}[WalkSAT lower bound]
\label{prop:direct-ws-lower}
Define
\[
  \gamma=\frac18\log_2\left(\frac1\rho\right),
  \qquad L=\left\lfloor2^{\gamma|V|}\right\rfloor.
\]
Under independent uniform initialization, for all sufficiently large $|V|$,
\begin{equation}
  \Prob(T_{\WS}\ge L)
  \ge1-\exp(-|V|/2)-\frac{486}{451}2^{-\gamma|V|}.
  \label{eq:direct-ws-tail-lower}
\end{equation}
Consequently, $\E{T_{\WS}}=2^{\Omega(|V|)}$.
\end{proposition}

\begin{proof}
Under uniform initialization, the blocks are independently active with
probability $1/4$.  Hence $|S_0|\sim\operatorname{Bin}(|V|,1/4)$, and
Hoeffding's inequality gives
\begin{equation}
  \Prob(|S_0|\ge H)
  \le\Prob\left(|S_0|-\frac{|V|}{4}\ge\frac{|V|}{2}\right)
  \le\exp(-|V|/2).
  \label{eq:direct-ws-initial-barrier}
\end{equation}

Each step changes $|S_t|$ by at most one.  Thus, on $\{|S_0|<H\}$, a
trajectory reaching $S_t=V$ must first move from $|S_t|=H-1$ to
$|S_t|=H$.  Each such entrance begins an attempt that ends at the next visit
to $H-1$ or $|V|$.  Conditional on the full history at the entrance,
Proposition~\ref{prop:direct-ws-one-attempt} bounds its success probability by
$\rho^{|V|-H}$.

At most one attempt begins in each step.  Thus a conditional union bound and
\eqref{eq:direct-ws-initial-barrier} give, for every $t\ge0$,
\begin{equation}
  \Prob(T_{\WS}<t)
  \le\exp(-|V|/2)+t\rho^{|V|-H}.
  \label{eq:direct-ws-attempt-union}
\end{equation}
Since $H\le3|V|/4+1$, the definition of $\gamma$ gives
\[
  L\rho^{|V|-H}
  \le2^{\gamma|V|}\rho^{|V|/4-1}
  =\frac{486}{451}2^{-\gamma|V|}.
\]
Substituting $t=L$ into \eqref{eq:direct-ws-attempt-union} proves
\eqref{eq:direct-ws-tail-lower}.  Since $\rho<1$, we have $\gamma>0$, and both
error terms on its right-hand side tend to zero.  Hence, for all sufficiently
large $|V|$,
\[
  \Prob(T_{\WS}\ge L)\ge\frac12.
\]
For the same values of $|V|$,
\[
  L=\left\lfloor2^{\gamma|V|}\right\rfloor
  \ge2^{\gamma|V|-1}.
\]
Pointwise,
$T_{\WS}\ge L\mathbf{1}_{\{T_{\WS}\ge L\}}$.  Taking expectations therefore
gives
\[
  \E{T_{\WS}}
  \ge L\Prob(T_{\WS}\ge L)
  \ge 2^{\gamma|V|-1}\cdot\frac12
  =2^{\gamma|V|-2}.
\]
Because $\gamma$ is a positive constant, this proves
$\E{T_{\WS}}=2^{\Omega(|V|)}$.
\end{proof}

For completeness, a monotone sequence of favorable flips gives a matching
coarse exponential upper bound.

\begin{proposition}[Exponential WalkSAT upper bound]
\label{prop:direct-ws-upper}
For every initial assignment $X_0$,
\begin{equation}
  \E{T_{\WS}\mid X_0}
  \le2|V|\left(\frac{2763}{52}\right)^{2|V|}.
  \label{eq:direct-ws-exp-upper}
\end{equation}
In particular, $\E{T_{\WS}}=2^{O(|V|)}$.
\end{proposition}

\begin{proof}
At every time $t$ before termination,
Lemma~\ref{lem:direct-structural-audit} and $102$-regularity give
\[
  B(X_t)\le921|Z_{0,t}|+154|Z_{1,t}|
  \le921(|Z_{0,t}|+|Z_{1,t}|).
\]
The number of unsatisfied clauses from four-clause gadgets is
\[
  104|Z_{0,t}|+52|Z_{1,t}|
  \ge52(|Z_{0,t}|+|Z_{1,t}|).
\]

After one of these clauses is selected, WalkSAT flips its zero progress
bit with probability $1/3$.  Therefore, conditional on the complete history,
the next step is such a favorable flip with probability at least
\[
  \eta=\frac{52}{3\cdot921}=\frac{52}{2763}.
\]
A favorable flip changes one zero progress bit to one and no other progress
bit.  Since there are at most $2|V|$ zero progress bits, consecutive favorable
flips reach $S_t=V$ within at most $2|V|$ steps.

Conditional on the history at the start of any interval of $2|V|$ steps, the
probability of termination during that interval is at least $\eta^{2|V|}$.
Applying this estimate successively to consecutive intervals and summing the
resulting geometric tail gives
\[
  \E{T_{\WS}\mid X_0}
  \le\frac{2|V|}{\eta^{2|V|}}
  =2|V|\left(\frac{2763}{52}\right)^{2|V|},
\]
which proves \eqref{eq:direct-ws-exp-upper}.
\end{proof}

\section{Conclusion}
\label{sec:conclusion}

In this paper, we have characterized a sharp occurrence threshold in the worst-case expected running times of WalkSAT and the MT algorithm on bounded-occurrence 3-CNF. We showed that, when every variable occurs at most twice, both algorithms find a satisfying assignment in expected linear time. This guarantee is tight with respect to the maximum variable-occurrence bound: once variables are allowed to occur three times, there exist satisfiable 3-CNF instances on which both WalkSAT and the MT algorithm require superpolynomial expected time. Thus, the worst-case behavior of both algorithms undergoes a sharp transition between maximum occurrence two and maximum occurrence three.

This dynamical transition is notably different from the classical satisfiability transition for bounded-occurrence 3-CNF. It is known that every 3-CNF formula in which each variable occurs at most three times is satisfiable, whereas deciding satisfiability becomes NP-complete when variables may occur up to four times~\cite{tovey1984simplified}. Moreover, in the occurrence-three regime, a satisfying assignment can be constructed in polynomial time through a matching argument. Consequently, the threshold at which WalkSAT and MT lose their universal polynomial expected-time guarantees occurs strictly below the threshold at which the underlying satisfiability problem becomes computationally hard. This separation exposes a fundamental limitation of both local-repair algorithms: even when satisfiability is guaranteed and a satisfying assignment
can be found efficiently by a global combinatorial method, their
local stochastic dynamics may still require superpolynomially
many updates in expectation.

We further established that WalkSAT and the MT algorithm are polynomially incomparable on 3-CNF. Although both algorithms repeatedly select an unsatisfied clause and attempt to repair it locally, their update rules lead to fundamentally different behaviors. We constructed a family of proper 3-CNF formulas on which WalkSAT runs in expected polynomial time whereas MT requires exponential expected time, as well as another family on which MT runs in expected linear time whereas WalkSAT requires exponential expected time. Therefore, under the initialization and selection rules studied in this paper, neither algorithm polynomially dominates the other, and neither provides a universal polynomial-overhead simulation of the other at the level of expected running time. Equivalently, the classes of 3-CNF instances efficiently handled by the two algorithms are incomparable.

Taken together, these results delineate distinct algorithmic capability boundaries for single-variable-flip local search and clause-wide resampling. They show that the apparent similarity between WalkSAT and MT is only superficial: changing the local repair operator can fundamentally alter the induced stochastic process and, consequently, the time required to reach a satisfying assignment. A natural direction for future work is to identify structural properties of 3-CNF instances that predict which of these two repair mechanisms is more effective, thereby developing finer criteria than the maximum variable-occurrence bound alone.

\section*{AI methodology}
\addcontentsline{toc}{section}{AI methodology}
We used ChatGPT (GPT-5.4, GPT-5.5, GPT-5.6, and GPT-6 Pro), Gemini 3.1 Pro (Deep Research), GPT-5.6 (Sol), and GPT-6 (Astra) to assist in constructing the clauses and analyzing the running times of the two algorithms. These tools materially affected Sections 3–6. The authors independently verified the correctness and originality of all content, including references.


\bibliographystyle{alphaurl}
\bibliography{ref}

\end{document}